\documentclass[final,5p,times,twocolumn]{elsarticle}

\usepackage{lineno}
\modulolinenumbers[5]

\usepackage[colorlinks,
linkcolor=black,
anchorcolor=black,
citecolor=black
]{hyperref}

\usepackage{threeparttable}
\usepackage{amssymb}
\usepackage{algorithm}
\usepackage[noend]{algorithmic}
\usepackage{epsfig}
\usepackage{subfigure}
\usepackage{multirow}
\usepackage{float}
\usepackage{amsmath}
\usepackage{longtable}
\usepackage{booktabs}
\usepackage{color}
\usepackage{subfigure}
\usepackage{fancyhdr}
\usepackage{epstopdf}
\usepackage{ulem}
\usepackage{bm}
\usepackage{amsopn}
\usepackage{times}
\usepackage{lineno}

\usepackage{stfloats} 

\journal{**}

\begin{document}

\begin{frontmatter}



\title{Task Graph offloading via Deep Reinforcement Learning in Mobile Edge Computing}


%

\author[lab1]{Jiagang Liu}
\ead{jiagliu@hnit.edu.cn}
\author[lab1]{Yun Mi \corref{mycorrespondingauthor}}
\cortext[mycorrespondingauthor]{Corresponding authors}
\ead{15111572562@163.com}
\author[lab2]{Xinyu Zhang\corref{mycorrespondingauthor}}
\ead{zhangxinyu@whu.edu.cn}
\author[lab2]{Xiaocui Li}
\address[lab1]{School of Computer Science and Engineering, Hunan Institute of Technology, Hengyang, Hunan}
\address[lab2]{School of Computer Science, Hunan University of Technology and Business, Changsha, China}

\begin{abstract}
Various mobile applications that comprise dependent tasks are gaining widespread popularity and are increasingly complex. These applications often have low-latency requirements, resulting in a significant surge in demand for computing resources. With the emergence of mobile edge computing (MEC), it becomes the most significant issue to offload the application tasks onto small-scale devices deployed at the edge of the mobile network for obtaining a high-quality user experience. However, since the environment of MEC is dynamic, most existing works focusing on task graph offloading, which rely heavily on expert knowledge or accurate analytical models, fail to fully adapt to such environmental changes, resulting in the reduction of user experience. This paper investigates the task graph offloading in MEC, considering the time-varying computation capabilities of edge computing devices. To adapt to environmental changes, we model the task graph scheduling for computation offloading as a Markov Decision Process (MDP). Then, we design a deep reinforcement learning algorithm (SATA-DRL) to learn the task scheduling strategy from the interaction with the environment, to improve user experience. Extensive simulations validate that SATA-DRL is superior to existing strategies in terms of reducing average makespan and deadline violation.

\end{abstract}




\end{frontmatter}


\section{Introduction}
With the rapid advance of wireless communication technologies and the proliferation of smart mobile devices, such as smartphones, tablets, and wearable devices, these lightweight smart devices have become significant terminals for mobile users (MUs) to access the Internet. \textcolor{blue}{Various mobile applications are gaining widespread popularity and are increasingly complex~\cite{JIT2217}.} Moreover, they are often composed of dependent tasks\cite{geng2018energy} that form the construct of a task graph, such as gesture recognition, mobile healthcare, and augment reality\cite{novak2018ultrasound,goudarzi2020application}. These applications often have low-latency requirements, resulting in a significant surge in demand for computing resources. Nevertheless, it is difficult for smart devices to effectually support the local execution of these computation-intensive applications due to the naturally lightweight and limited computing resources\cite{ren2019survey}. The tension between computation-intensive applications and resource-limited smart devices generates a bottleneck for obtaining a high-quality user experience. 

The emergent mobile edge computing (MEC) is proposed to deploy many small-scale devices, such as micro base stations, edge routers, and roadside units, at the edge of the mobile network in close proximity to users, instead of the remote cloud. Edge computing devices (ECDs) can provide richer computation resources for MUs through wireless access service with reliable low-latency communications\cite{LIU2022228,liu2022auction}. With computation offloading technology, the tasks of computation-intensive applications may be migrated to ECDs, not via the core network, so that the processing of these applications can be expedited, resulting in a high-quality user experience. As different tasks in mobile applications require different workloads and the amount of transfer data, the task scheduling strategy for computation offloading needs to determine which tasks are processed on which ECDs and which tasks are executed locally. As a result, the task offloading strategy becomes the most significant issue in MEC\cite{duan2022distributed}.

Most works have focused on optimal task scheduling strategies for task graph offloading in terms of the application completion time\cite{zhang2018resource}, energy consumption\cite{sundar2018offloading}, and communication service\cite{naouri2021novel}. Due to the NP-hardness of task scheduling optimization, most studies on task graph offloading in MEC are based mainly on heuristic or approximation methods\cite{sundar2018offloading,zhao2020offloading,al2020task,meng2016approach}. \textcolor{blue}{In real-world scenarios, ECDs in the network are mainly responsible for some specialized functions\cite{dong2022wave,IOT9320569}, such as routing, relaying, forwarding, and so on.} In this case, ECDs only generally share their idle computing resources with the applications offloaded from MUs. Considering that ECDs are still relatively resource-constrained devices compared to the remote cloud, their idle computing resources are incapable of maintaining persistent performance. The computing capabilities of ECDs may vary with time making the MEC environment dynamic. However, these aforementioned works rely heavily on expert knowledge or accurate analytical models so they fail to fully adapt to the dynamic environment, resulting in the deadline violation for the mobile applications since the actual task execution may fluctuate frequently\cite{liu2019online,al2020task,zhao2020offloading,wang2021dependent}. Consequently, a few works have been attracted to research the computation offloading in MEC by utilizing the novelty technologies that can adapt to the dynamic environment.

Deep reinforcement learning (DRL), which combines reinforcement learning (RL) with a deep neural network, is a promising method that adaptively and flexibly makes sequential decisions in a dynamic environment, without the need for expert knowledge. As a result, some researchers have begun to focus on how to solve the problem of computational offloading in dynamic MEC with DRL.

A few works investigate the task offloading in MEC by utilizing DRL to adapt to the environment variation\cite{he2017integrated,ning2019deep,chen2018optimized,xu2019confidence}. \textcolor{blue}{These works consider the dynamic variation of the edge computing network to improve the performance of computation offloading. However, these works focus the coarse-grained task offloading rather than fine-grained task graph offloading\cite{dinh2017offloading}. Coarse-grained task offloading regards an application as an indivisible whole and designs the offloading scheme based on applications' requirements on computing resources. This scheme does not divide the mobile application into tasks from its functional point of view and may result in unexpected computation delay by unreasonable processing compositor of tasks with dependency constraints.}

\textcolor{blue}{Recently, some works apply DRL to investigate the fine-grained offloading in MEC\cite{song2022offloading,wang2021dependent}. Song et al.~\cite{song2022offloading} took into account the different dynamic preferences between MEC and users to solve the multi-objective optimization problem of task graph offloading in MEC. Wang~\cite{wang2021dependent} focused on the variation of dependent constraints between tasks to improve the Quality-of-Service (QoS) of task graphs. Although these works have proposed various task graph offloading strategies, they did not consider the variation of computing capabilities of ECDs. The dynamic variation will lead to task execution delays. Moreover, the effect will continuously accumulate until task graph applications are complete if the offloading strategy can not adapt to the dynamic variation, which leads to the deadline violation. This consequently motivates us to investigate DRL-based task graph offloading to improve user experience, considering the time-varying computing capabilities of ECDs.}

\textcolor{blue}{In this paper, we propose a task graph offloading strategy based on DRL in the MEC system. The proposed strategy can adapt to the dynamic environment when the computing capabilities of ECDs are time-varying, to satisfy application deadlines and improve user experience as much as possible. We first formulate the task graph offloading in MEC as an optimization problem of minimizing the average application completion. Based on the topological ordering of applications, we decompose the ready tasks of the arrival applications into a prioritized list. Then, we model innovatively the scheduling processing of the ready tasks as a Markov Decision Process (MDP). In this MDP, the characterization of the environment is formulated as the state space, and the scheduling decision for each ready task is abstracted into the action space. The reward with respect to MDP is defined as the benefit for the agent. To overcome the difficulties caused by the large number of state spaces, we designed a DRL algorithm based on DQN to learn the scheduling decision from the interaction with the environment. Particularly, the major contributions of this paper are summarized as follows.}


\begin{itemize}
	\item \textcolor{blue}{The task graph offloading is decomposed into making the scheduling decision for the ready tasks of the arrival applications in a sequence of discrete time steps. The scheduling processing of the ready tasks is modeled as an MDP innovatively.}
	
	\item \textcolor{blue}{The environmental factors to be considered when scheduling each task schedule are formulated as the state space, and the scheduling decision for each task is abstracted into the action space. A reward value is formulated with novelty, consisting of a utility function, a duration factor, and a penalty factor.}

	\item \textcolor{blue}{A DRL algorithm based on DQN is designed to learn the scheduling decision from the interaction with the environment. By conducting extensive simulations, the proposed task graph offloading algorithm is superior to existing strategies, in terms of reducing average makespan and deadline violation.}
\end{itemize}

The remainder of this paper is organized as follows. Section~\ref{related} \textcolor{blue}{briefly surveys the related work.} Section~\ref{SystemModel} presents the system model and Section~\ref{Formulation} formulates the optimization problem of the task graph offloading in MEC. Section~\ref{Mechanism} gives the RL-based scheduling mechanism, and Section~\ref{ResourceAllocation} describes the resource allocation based on RL. Then, the simulation evaluation and conclusions are provided in Section~\ref{Performance} and Section~\ref{Conclusion}, respectively.

\section{Related work}
\label{related}
\textcolor{blue}{With advance of the Internet of Things (IoT) and Big data, smart mobile applications and IoT applications have proliferated rapidly~\cite{IOT8930011,JIT2378}. Increasing researchers pay attention to the computing offloading in the edge computing. We briefly survey the related work in this section.}

\textcolor{blue}{Task graph scheduling has always been an important issue in the heterogeneous computing systems~\cite{li2013energy,xu2014genetic}. Some works extended the traditional task scheduling techniques in cloud computing to the edge computing. Zhang et al.~\cite{zhang2018resource} extended the partial critical path (PCP) to allocate computing resources for the task graph application. Our previous work~\cite{liu2022auction} regarded the computing resource allocation based on PCP for the task graph application as a trade. Then, we proposed an auction-based game to scheduling task graph applications when satisfying deadline constraints. Nevertheless, these works did not consider the MEC environment dynamic.}

\textcolor{blue}{To adapt to the dynamic environment, many researchers have applied DRL to the computation offloading in dynamic MEC. He et al.~\cite{he2017integrated} improved the performance of computation offloading for vehicular networks by dynamic orchestration of network, caching, and computing resources. Ning et al.~\cite{ning2019deep} considered the variation of the channel state and the computation capability and proposed a DRL scheme to optimize traffic scheduling and resource allocation in vehicular networks. Chen et al.~\cite{chen2018optimized} took into account the time-varying network dynamics to investigate an optimal computation offloading policy in a sliced radio access network. They proposed a DRL-based offloading decision to maximize the long-term utility performance. However, these works did not take into account the dependency among tasks, that is, they focused on the coarse-grained task offloading.}

\textcolor{blue}{There are also a few advances in utilizing DRL to further optimize the computing resource allocation of task graphs in the MEC system. Lu et al.~\cite{lu2020optimization} focused on the scenarios of the multi-service nodes and task graph applications in heterogeneous MEC. They proposed a fine-grained offloading scheme based on DRL to achieve the reduction of execution latency, monetary cost, energy consumption, as well as network usage. Song et al.~\cite{song2022offloading} considered the different dynamic preferences between MEC and users to investigate the offloading decision. They proposed a multi-objective optimization strategy based on DRL for task graph applications in MEC. Wang et al.\cite{wang2021dependent} proposed a DRL-based task graph offloading scheme to make the task offloading plan, considering the variation of dependent constraints between tasks. These above-mentioned works did not take into account the variational computing capabilities of ECDs, though they improved the performance of task graphs.}

\textcolor{blue}{Different from all the existing works above, this work investigates the fine-grained task graph offloading and proposes an offloading strategy based on DRL to adapt to the variational computing capabilities of ECDs.}

\begin{figure}[htbp]
	\centering
	\includegraphics[width=0.45\textwidth]{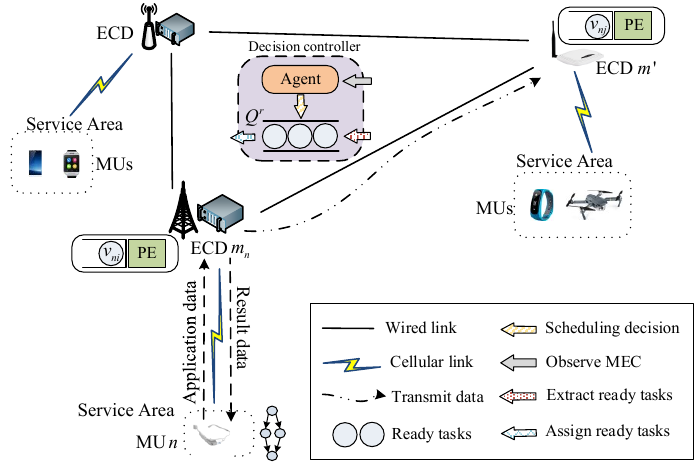}
	\caption{\textcolor{blue}{Components of MEC system.}}
	\label{MEC_System}
\end{figure}

\section{System Mobel}
\label{SystemModel}
A MEC system, as shown in Fig.~\ref{MEC_System}, is comprised of many heterogeneous ECDs deployed at the edge of the mobile network. {\color{blue}There exists a decision controller that resides in some ECDs in the system~\cite{zhang2017resource}. It can periodically collect the real-time status information broadcasted from all ECDs. In the decision controller, we design an intelligent agent of the reinforcement learning, unlike traditional task schedulers (e.g., YARN). The agent makes the decision intelligently for the task scheduling by observing changes in the MEC system.} 

Let $\mathcal{M}$ express the set of ECDs in the MEC system and $M=|\mathcal{M}|$ denote the total number of ECDs. Then, we use $m$ ($\forall m\in\{1,\ldots, M\}$) to index the $m$-th ECD. These ECDs in $\mathcal{M}$ communicate with each other via wired links~\cite{zhang2017resource}. We indicate the transmission rate between any two ECDs as $B_{m,m'}$, where $m\neq m'$ and $\forall m,m'\in \{1,\ldots,M\}$. Based on the literature~\cite{zhang2015energy}, $B_{m,m'}$ can be estimated generally. Due to physical constraints, the computing resources of ECDs are limited. We assume that each ECD has only one processing element (PE) which provides the computing service for the task execution of the application following First Come First Service manner (FCFS). In practice, the processing capabilities of PEs are finite continuous values varying with time~\cite{ning2019deep}. Moreover, the processing capability state at the next moment is related to the previous moment. We assume that the processing capabilities of PEs are stabilized over a short period when they are within a level. Therefore, the processing capabilities of PEs can be discretized and quantized into several levels, similar to the literature~\cite{ning2019deep}. Then, let $\delta_m$ express the processing capability of ECD $m$ at a state, which can be measured in millions of instructions per second (MIPS). {\color{blue}Each ECD can periodically broadcast the current status information, such as its own workload and the transmission rate with other ECDs. Since the work mainly focuses designing an intelligent algorithm to allocate the dependent tasks among different connected ECDs, we did not pay a special focus on the task scheduling over parallel processing elements. This simple model can also be easily extended to a more general case in which each ECD has multiple PEs with parallel computing capability by adding more virtual ECDs and neglecting the data transmission cost among them.}

In the MEC system shown in Fig.~\ref{MEC_System} each ECD can provide cellular communication services to many MUs in a specified area. We use $\mathcal{N}$ to indicate the set of all MUs that the MEC system can serve, and $N=|\mathcal{N}|$ denotes the total number of MUs. Thus, $n$ ($\forall n\in\{1,\ldots, N\}$) can index the $n$-th MU. Each MU needs to process an application composed of multiple dependent tasks. This proposal can also be easily extended to the scenarios in which a MU can process multiple applications by adding some new MUs that can only process one application to the system. As a result, $\mathcal{N}$ also represents the set of all applications and $n$ can index the $n$-th application\footnote{In this paper, we will interchangeably use three terms, i.e., MU $n$, $n$-th MU and application $n$, to represent a user application}. We can model the application as a direct acyclic graph (DAG)~\cite{sundar2018offloading}. Then, $\mathcal{G}_n(\mathcal{V}_n,\mathcal{E}_n)$ indicates a DAG corresponding to application $n$. $\mathcal{V}_n$ denotes the set of tasks. $\mathcal{E}_n$ is the set of data dependencies, called directed edges, among these tasks. For any task in $\mathcal{V}_n$, it is an atomic and indivisible component.

\begin{figure}[!t]
	\centering
	\includegraphics[width=0.48\textwidth]{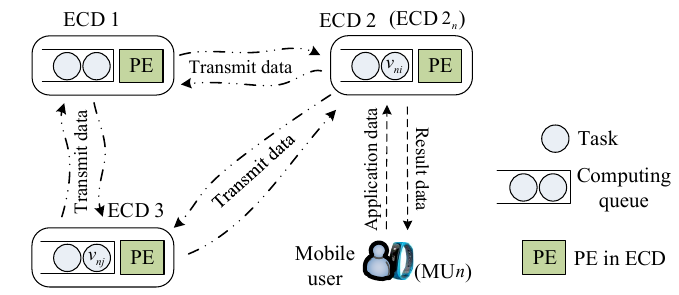}
	\caption{Network Topology.}
	\label{Topology}
\end{figure}

MU $n$ expects to expedite the execution of its application before the deadline, so it needs to offload application data to the MEC system. To achieve this, MU $n$ must first send the offloading request to the ECD whose cellular communication service covers MU $n$. Let $m_n$ indicate the ECD that covers MU $n$, as shown in Fig.~\ref{MEC_System}. $B_n^m$ denotes the transmission rate between ECD $m_n$ and MU $n$. After ECD $m_n$ accepts this request and receives the application data sent by MU $n$, the decision controller will extract the tasks ready for execution from ECD $m_n$ and push them into a ready queue $Q^r$. Whereafter, the decision controller can make the assignment for the tasks in $Q^r$ according to a scheduling strategy. Moreover, any task can only be executed on one ECD. The result data must send back to MU $n$ after all tasks of application $n$ are completed in the MEC system. As MU $n$ is located within the cellular signal coverage area of ECD $m_n$, the result data needs to be transmitted to MU $n$ via ECD $m_n$. For any MU $n$, it has the same network topology for task execution as all ECDs in this system can execute tasks offloaded from MU $n$. Hence, we only focus on the topology with one MU for simplicity, which is shown in Fig.~\ref{Topology}. More MUs connected to the MEC system only increase the number of applications offloaded to this system, which does not affect this study. Built upon this topology with one MU, we can assume MU $n$ as the fictitious ECD which is the $0$-th edge computing device in $\mathcal{M}$. To model the application data sent by MU $n$ and the result data received by MU $n$, we add two dummy tasks to each application in $\mathcal{N}$. Moreover, the two dummy tasks must be on this fictitious ECD corresponding to MU $n$. As a result, all tasks in application $n$ can be offloaded to the MEC system.

Considering the above characteristics of the computing offloading, we use the triple elements $\{r_n, d_n, \mathcal{G}_n\}$ to indicate the application that MU $n$ offloads to the MEC system. Here, $r_n$ denotes the time of the offloading request sent by MU $n$ and $d_n$ is the deadline of application $n$.  In the task graph $\mathcal{G}_n$, $I_n=|\mathcal{V}_n|$ expresses the number of tasks in application $n$. We indicate each task in $\mathcal{V}_n$ as  $v_{ni}$ ($i\in\{0,\ldots, I_n\}$). Here, $i$ indexes the $i$-th task. For ease of expression, meanwhile, we use $v_{n0}$ and $v_{nI}$ to denote the two dummy tasks, respectively. For task $v_{ni}$, it has a workload $\rho_{ni}$, which is the number of instructions needed by the PE to process $v_{ni}$. Generally, $\rho_{ni}$ can be acquired by profile analytics~\cite{venkataraman2016ernest}. Accordingly, $\rho_{n0}$ and $\rho_{nI}$ are set to zero since $v_{n0}$ and $v_{nI}$ are the dummy tasks added to application $n$. 
{\color{blue}Due to the dependency constraints among tasks, the output data from task $v_{ni}$ needs to be transferred to another task $v_{nj}$. Notice that, the dummy $v_{nI}$ has no the output data and the dummy $v_{n0}$ has no the input data. The data transfer between $v_{ni}$ and $v_{nj}$ forms a directed edge, denoted as $\varepsilon_{nij}\in \mathcal{E}_n$. If there exists the directed edge from $v_{ni}$ to $v_{nj}$, we call $v_{ni}$ the parent of $v_{nj}$. Accordingly, $v_{nj}$ is the child of $v_{ni}$.} We use $p(v_{ni})$ and $c(v_{ni})$ to define the sets of $v_{ni}$'s parents and children, respectively. For the two dummy $v_{n0}$ and $v_{nI}$, especially, $p(v_{n0})$ and $c(v_{nI})$ are empty sets. Under the scheduling strategy made by the decision controller, the output data from a completed task need to be transferred to the ECD where the child tasks are located, due to the data dependencies between tasks. For example, task $v_{ni}$ is on ECD $2_n$ and $v_{ni}$'s child $v_{nj}$ is on ECD $3$ in Fig.~\ref{Topology}. After ECD $2_n$ completes task $v_{ni}$, ECD $2_n$ needs to transmit the all output data to ECD $3$ via the communication link since $v_{ni}$'s child $v_{nj}$ is on ECD $3$. Let $e_{nij}$ denote the size of data transfer from $v_{ni}$ to $v_{nj}$. Especially, if both $v_{ni}$ and its child $v_{nj}$ on the same ECD $m$, the transmit data $e_{nij}$ from $v_{ni}$ can be delivered directly to $v_{nj}$ without going through the communication link. {\color{blue}To help understand, the common symbols used throughout this paper are listed in Table~\ref{Symbols}.}

\begin{table}[htbp]
	\centering
	\caption{Frequently Used Symbols}  
	\begin{tabular}{p{1.5cm}p{6.6cm}} 
		\toprule  
        Symbol&  Definition\\  
		\midrule 
		$M$, $n$ & \textcolor{blue}{The total number of ECDs and the number of MUs}\\
		$B_{m,m'}$ & \textcolor{blue}{The transmission rate between $m$ and $m'$}\\
		$m_n$ & \textcolor{blue}{The ECD covering MU $n$}\\
		$B_n^m$ & \textcolor{blue}{The transmission rate between ECD $m_n$ and MU $n$}\\
		$\hat{B}$ & \textcolor{blue}{The maximum value of all $B_{m,m'}$}\\
		$\delta_m$ & \textcolor{blue}{The processing capability of ECD $m$}\\
		$\hat{\delta}$ & \textcolor{blue}{The maximum value of all $\delta_m$}\\		
		$\mathcal{V}_n$, $\mathcal{E}_n$ & \textcolor{blue}{The sets of tasks and directed edges in DAG $\mathcal{G}_n$}\\
		$I_n$ & \textcolor{blue}{The number of tasks in application $n$}\\
		$v_{ni}$ & \textcolor{blue}{The $i$-th task in application $n$}\\
		$v_{n0}$ and $v_{nI}$ & \textcolor{blue}{Two dummy tasks in application $n$}\\
	    $\rho_{ni}$ & \textcolor{blue}{A workload of task $v_{ni}$}\\
	    $\varepsilon_{nij}$ & \textcolor{blue}{The directed edge between $v_{ni}$ and $v_{nj}$}\\
	    $p(v_{ni})$ & \textcolor{blue}{The set of $v_{ni}$'s parents}\\
	    $c(v_{ni})$ & \textcolor{blue}{The set of $v_{ni}$'s children}\\
	    $\textbf{x}_{ni}$ & \textcolor{blue}{A strategy for task $v_{ni}$}\\
	    $x_{ni}^m$ & \textcolor{blue}{A binary variable denoting the scheduling plan}\\
	    $E(\textbf{x}_{ni})$ & \textcolor{blue}{The execution time of $v_{ni}$}\\
	    $e_{nij}$ & \textcolor{blue}{The transmit data between $v_{ni}$ to $v_{nj}$}\\
	    $T(\textbf{x}_{ni},\textbf{x}_{nj})$ & \textcolor{blue}{The data transfer time from $v_{ni}$ to $v_{nj}$ via the communication link}\\
	    $F(\textbf{x}_{ni})$ & \textcolor{blue}{The completion time of $v_{ni}$}\\
	    $A(\textbf{x}_{nj})$ & \textcolor{blue}{The arrival time of input data to $v_{ni}$}\\
	    $F^{lct}_{ni}$ & \textcolor{blue}{The latest completion time of $v_{ni}$}\\
	    $\xi_n$ & \textcolor{blue}{The list of tasks in application $n$ in ascending order}\\
	    $\textbf{s}_\tau$ & \textcolor{blue}{The state space in MDP at time step $\tau$}\\
	    $\textbf{a}_\tau$ & \textcolor{blue}{The action space in MDP at time step $\tau$}\\
	    $\textbf{r}$ & \textcolor{blue}{The reward in MDP}\\ 
		\bottomrule  
	\end{tabular}
	\label{Symbols}   
\end{table}

\begin{figure}[!t]
	\centering
	\includegraphics[width=0.35\textwidth]{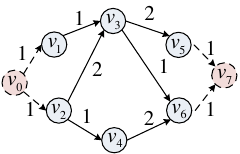}
	\caption{Task Graph Model.}
	\label{DAG}
\end{figure}

To ease the understanding, Fig.~\ref{DAG} shows an example of the task graph model with eight tasks. This task graph for an application contains eight tasks, where the dotted red circles indicate dummy tasks and the gray circles denote real tasks. The dotted arrows express the application data to be offloaded and the result data to be sent back to MU, respectively. Specifically, the former are the outgoing arcs attached to $v_0$ and the latter correspond to the incoming arcs attached to $v_7$. Moreover, the numbers attached to the directed arcs, i.e., solid arrows and dotted arrows, denote the size of transfer data. 

\section{Problem Formulation}
\label{Formulation}
We define a strategy $\textbf{x}_{ni}=(x_{ni}^0,\ldots,x_{ni}^M)$ for task $v_{ni}$ in application $n$. Peculiarly, the item $x_{ni}^m$ ($\forall m\in \{0,\ldots,M\}$) in $\textbf{x}_{ni}$ is the binary variable that denotes the scheduling plan. Since each real task $v_{ni}$ ($i\notin \{0,I\}$) can be executed on any real ECD $m$ ($m\neq 0$), we can define $x_{ni}^m$ as follows.
\begin{eqnarray}
	\begin{aligned}
		\label{Binary}
		x_{ni}^m=\begin{cases}1, & \text{if ECD }m\text{ can execute }v_{ni},\\
			0, &\text{otherwise}.\end{cases}
	\end{aligned}
\end{eqnarray}
For the dummy tasks $v_{n0}$ and $v_{nI}$, they must be on the ECD $0$ corresponding to MU $n$. Hence, we have
\begin{equation}
	\label{dummy}
	x_{n0}^0=1,\quad x_{nI}^0=1.
\end{equation}
For the other real tasks $v_{ni}$, i.e., $i\neq\{0,I\}$, they need to be offloaded to the MEC system rather than executed locally. Thus, we have
\begin{equation}
	\label{real}
	x_{ni}^0=0, \quad i\neq \{0,I\}.
\end{equation}
As any task can only be scheduled to one ECD for execution, we have
\begin{equation}
	\label{OneAssign}
	\sum\limits_{m= 0}^{M}{x_{ni}^m=1}, \quad \forall i\in\{0,\ldots,I\}.
\end{equation}
If task $v_{ni}$ is scheduled to ECD $m$ under the strategy $\textbf{x}_{ni}$, we can define the execution time $E(\textbf{x}_{ni})$ of $v_{ni}$ on ECD $m$ as follows.
\begin{equation}
	\begin{aligned}
		\label{ExeTime}
		E(\textbf{x}_{ni})=\begin{cases} 0, &\text{if }i\in\{0,I\},\\
			\sum\limits_{m= 1}^{M}{\frac{\rho_{ni}}{\delta_m}\cdot x_{ni}^m}, &\text{otherwise}.
		\end{cases}
	\end{aligned}	
\end{equation}

Since there exists transmit data $e_{nij}$ between two tasks $v_{ni}$ and $v_{nj}$ with the dependency constraints, we use $T(\textbf{x}_{ni},\textbf{x}_{nj})$ to indicate the data transfer time via the communication link. Considering that $e_{nij}$ from $v_{ni}$ can be delivered to $v_{nj}$ without going through the communication link when $v_{ni}$ and $v_{nj}$ are scheduled to the same ECD, we define $T(\textbf{x}_{ni},\textbf{x}_{nj})=0$ when $\textbf{x}_{ni}=\textbf{x}_{nj}$. If $v_{ni}$ is the task to which the application data offloaded from MU $m$ is to be transferred, i.e., $v_{n0}\in p(v_{ni})$, there are two cases of ECD $m$ that $v_{ni}$ can be scheduled to. In one case, $m=m_n$, that is, MU $n$ is in the cellular signal area of ECD $m$. Thus, MU $n$ can connect to ECD $m$ directly. Hence, we have
\begin{equation}
	\label{In_M_n}
	T(\textbf{x}_{n0},\textbf{x}_{ni})=\frac{e_{n0i}}{B_n^m}, \quad v_{ni}\in c(v_{n0}),\text{ and }m=m_n,
\end{equation}
where $B_n^m$ is the transmission rate between MU $n$ and ECD $m_n$, and $e_{n0i}$ is the size of data transfer from $v_{n0}$ to $v_{ni}$. In the other, $m\neq m_n$. The application data offloaded from MU $n$ must be transferred to ECD $m$ via ECD $m_n$. Therefore, we calculate $T(\textbf{x}_{n0},\textbf{x}_{ni})$ by
\begin{equation}
	\label{Via_M_n}
	\frac{e_{n0i}}{B_n^m}+\sum\limits_{m= 1}^{M}{\frac{e_{n0i}}{B_{m_n,m}}x_{ni}^m},\text{  }v_{ni}\in c(v_{n0}) \text{ and } m\neq m_n.
\end{equation}
Furthermore, if $v_{ni}$ is the task that outputs the result data to MU $n$, i.e., $v_{nI}\in c(v_{ni})$, there are also two cases of ECD $m$. For the first case, $m=m_n$, so that the result output by $v_{ni}$ on ECD $m$ can be transferred directly to MU $n$. Thus, we have
\begin{equation}
	\label{Out_M_n}
	T(\textbf{x}_{ni},\textbf{x}_{nI})=\frac{e_{niI}}{B_n^m}, \quad v_{ni}\in p(v_{nI}),\text{ and }m=m_n.
\end{equation}
In the second case, the result output by $v_{ni}$ on ECD $m$ must be transferred to MU $n$ via the ECD $m_n$ such that $T(\textbf{x}_{ni},\textbf{x}_{nI})$ can be defined as
\begin{equation}
	\label{Result_from}
	\frac{e_{niI}}{B_n^m}+\sum\limits_{m= 1}^{M}{\frac{e_{niI}}{B_{m,m_n}}x_{ni}^m},\text{ }v_{ni}\in p(v_{nI})\text{ and }m\neq m_n.
\end{equation}
Except for the aforementioned cases, $v_{ni}$ and $v_{nj}$ are respectively scheduled to real ECDs $m$ and $m'$ ($m\neq m'$) which can communicate directly with each other. We have
\begin{equation}
	\label{mesh}
	T(\textbf{x}_{ni},\textbf{x}_{nj})=\sum\limits_{m= 1}^{M}\sum\limits_{m'= 1}^{M}{\frac{e_{nij}}{B_{m,m'}}\cdot x_{ni}^m\cdot x_{nj}^{m'}},\quad m\neq m'.
\end{equation}

The decision controller can detect the completion time of task $v_{ni}$, after ECD $m$ completes $v_{ni}$. We use $F(\textbf{x}_{ni})$ to indicate $v_{ni}$'s completion time. If $v_{ni}$ is the dummy task $v_{n0}$, i.e., $i=0$, the time when MU $n$ sends the application data can be seen as the completion time of $v_{ni}$. If $v_{ni}$ is the dummy task $v_{nI}$, i.e., $i=I$, moreover, the time when MU $n$ receives the result data can be regarded as the completion time of $v_{ni}$. Therefore, we calculate $F(\textbf{x}_{ni})$ by
\begin{equation}
	\label{dummy_Finish}
	\begin{aligned}
		\begin{cases}r_n, &\text{if  }i=0,\\
			\mathop{\max}\limits_{v_{nj}\in p(v_{ni})}\{F(\textbf{x}_{nj})+T(\textbf{x}_{nj},\textbf{x}_{ni})\}+E(\textbf{x}_{ni}),&\text{if  }i=I.
		\end{cases}
	\end{aligned}	
\end{equation}
For any real task $v_{ni}$, it can be offloaded to ECD $m$ ($\forall m\in\{1,\ldots,M\}$) for execution. As ECD $n$ processes $v_{ni}$ following FCFS, we must take into account the queue delay on the ECD $n$ and the arrival of input data to $v_{ni}$. As a result, we can define $F(\textbf{x}_{ni})$ as follows.
\begin{eqnarray}
	\label{FinishTime}
	\begin{aligned}
		F(\textbf{x}_{ni})=\mathop{\max}\left\{Q(\textbf{x}_{ni}), \mathop{\max}\limits_{v_{nj}\in p(t_{ni})}A(\textbf{x}_{nj})\right\}+E(\textbf{x}_{ni}).
	\end{aligned}
\end{eqnarray}
In Eq.~(\ref{FinishTime}), $Q(\textbf{x}_{ni})$ expresses the time when ECD $n$ is ready to execute $v_{ni}$ under the scheduling plan $\textbf{x}_{ni}$. Specifically, $Q(\textbf{x}_{ni})=\sum\nolimits_{m= 1}^{M}Q_mx_{ni}^m$. Here $Q_m$ is the queue delay of ECD $m$, to which $v_{ni}$ will be scheduled. Moreover, $A(\textbf{x}_{nj})$ is the arrival time of input data to $v_{ni}$. We can define $A(\textbf{x}_{nj})$ as follows.
\begin{equation}
	\label{arrval}
	A(\textbf{x}_{nj})=F(\textbf{x}_{nj})+T(\textbf{x}_{nj},\textbf{x}_{ni}),\quad \forall v_{nj}\in p(v_{ni}).
\end{equation}
After all of MU $n$'s tasks are completed in the MEC system, the decision controller can obtain the completion time $\Psi_n$ of application $n$, i.e., the makespan. Therefore, we can calculate $\Psi_n$ based on the maximum completion time among all tasks. Hence, we have
\begin{equation}
	\label{makespan}
	\Psi_n= \mathop{\max}\limits_{v_{ni}\in \mathcal{V}_n}F(\textbf{x}_{ni}) -r_n.
\end{equation}

All MUs in the MEC system expect to expedite the processing of applications. Therefore, the objective of this paper is to design a scheduling strategy to minimize the average makespan of applications for all MUs. We can formulate the optimization problem as follows.
\begin{align}
	\label{OptObt1}
	\mathbf{P}:\quad\mathop{\min}_{\{\textbf{x}_{ni}\}}&\frac{1}{N}\sum\limits_{n= 1}^{N}\Psi_n\\
	\mbox{s.t.}\quad& (\ref{Binary}),(\ref{dummy}),(\ref{real}),(\ref{OneAssign}) \nonumber \\
	&x_{ni}^m\in\{0,1\},\nonumber \\ 
	&\forall n\in \mathcal{N}\quad\forall m\in \mathcal{M}\quad \forall i\in{1,\ldots,I}.
\end{align}
Constraint~(\ref{Binary}) denotes that the decision variable. Constraint~(\ref{dummy}) claims that the dummy tasks must be located in MU $n$, i.e., ECD $0$. Constraint~(\ref{real}) indicates that all real tasks are offloaded to the MEC system. Constraint~(\ref{OneAssign}) expresses that task $v_{ni}$ can only be scheduled to one ECD.

We can reduce Problem $\mathbf{P}$ to the multiprocessor scheduling problem which is a well-know NP-hard problem~\cite{lewis1983michael}. Moreover, it is more difficult to obtain the optimal solution 
as a result of the dynamic and variable nature of the network environment. Therefore, obtaining the optimal solution to Problem $\mathbf{P}$ will consume high time complexity. For any MU, it needs an efficient strategy with polynomial time rather than the optimal solution, due to the low latency requirement. {\color{blue}The heuristic or approximation methods can find a feasible solution to task offloading rapidly, but they rely heavily on expert knowledge or accurate analytical models~\cite{wang2021dependent}. It is also more difficult to formulate an accurate analytical model in a dynamic environment, which is time-consuming due to the increasing complexity of applications~\cite{wang2021dependent}. DRL can make sequential decisions by self-learning in a dynamic environment, without the need for expert knowledge.} In this work, we change the optimal control of Problem $\mathbf{P}$ into a Markov Decision Process (MDP) and introduce deep reinforcement learning to solve it.

\section{Reinforcement Learning-based Scheduling Mechanism for Task Graph}
\label{Mechanism}
Reinforcement learning is one of the methodologies that can make an optimal decision by self-learning in a specific environment. It can abstract an optimization problem into an interactive process between an agent and the environment. Reinforcement learning is suitable to make the optimal decision based on the reward signals received continuously from the state of the environment. In recent years, reinforcement learning has been widely applied to resource allocation and optimal control. Generally, the method of reinforcement learning can be modeled as a MDP. In this section, we consider the MEC system as an environment and model the task scheduling for applications of MUs as MDP. 

\subsection{Preparation for MDP}
To model a MDP, we must abstract the scheduling of the task graph into a sequence of discrete time steps. At each time step, the representation of the environment can be received by the proposed scheduling mechanism. In the following we present the modeling of discrete time steps for the task graph scheduling.

The agent to implement the reinforcement learning is resided in the decision controller. We define the operation of the agent as an event-driven mechanism. It will be motivated on receipt of two events: one is the arrival event, and the other is the completion event. Recall that MU $n$ needs to offload application data to the ECD, i.e., $m_n$, whose cellular communication service covers MU $n$. When the application data are offloaded to ECD $m_n$, the arrival event appears. After any task is completed by any ECD, the completion event triggers. Such two events will motivate the agent to work continuously. 

For each arrival application, the agent will confirm the task priorities. To achieve this, the list scheduling heuristic is used for the calculation of  task priorities. Since the decision controller in the MEC system is a centralized control unit, the agent can gather the processing capabilities of all ECDs in the system. Moreover, it can perceive the range of processing capability of each ECD. Built upon this, we can estimate the latest completion time of all tasks in an application and take it as the task priority by assuming that they are scheduled to the ECD with the maximum processing capability for parallel execution~\cite{kwok1999static}. The latest completion time $F^{lct}_{ni}$ of task $v_{ni}$ is estimated by
\begin{equation}
	\begin{aligned}
		\label{LCT}
		F^{lct}_{ni}=\begin{cases} d_n-e_{niI}/B_n^m, \quad\quad\quad\quad\quad\text{if } v_{ni}\in p(v_{nI}),\\ 
			\mathop{\min}\limits_{v_{nj}\in c(v_{ni})}\left\{F^{lct}_{nj}-\frac{\rho_{nj}}{\hat{\delta}}-\frac{e_{nij}}{\hat{B}}\right\},\text{otherwise},\end{cases}
	\end{aligned}
\end{equation}
where $\hat{\delta}$ and $\hat{B}$ are the maximum processing capability in all ECDs and the maximum transmission rate among all ECDs, respectively. The calculation of $\hat{\delta}$ and $\hat{B}$ does not take into account the devices of mobile uses. As the MEC system can serve multiple MUs, each arrived application needs to be calculated by the agent based on Eq.~(\ref{LCT}). Therefore, we construct a list  $\xi_n$ for each application $n$ arriving at different ECDs, in which the application tasks are sorted by the ascending order of the task priority. Obviously, the number of lists is equal to the number of applications to be processed concurrently by the system. We denote the set of the lists as $\mathcal{H}=\{\xi_n\}$~($\forall n\in \mathcal{N}$). 

As the execution of the task graph is constrained by the dependencies among tasks, the scheduling plan of a task can be made after its parents have been made. We call such a task the ready task. When the agent receives the arrival event or the completion event, it will pick out the ready tasks from $\xi_n$ and arrange them into the queue $Q^r$  in an ascending order of their priorities. Subsequently, the agent will fetch the first ready task from $Q^r$, in turn, and make the scheduling plan for it. Built upon such a mechanism, we think of the time when the agent gets each task out of $Q^r$ at a time step. The agent will interact with the environment to plan for task scheduling via a reinforcement learning algorithm at each time step.

\subsection{MDP Mode of Agent}
The MDP mode requires that the agent can receive reward signals by observing the state of the environment. The decision controller can periodically collect real-time status information about the system so that the agent can receive reward signals from the state of the environment. The reward signals are used to evaluate the quality of actions available to the agent. Then, the agent selects an action based on the evaluation. Afterward, at the next time step the environment changes into a new state. The agent can receive new reward signals according to the feedback of the environment. The agent chooses actions with the goal of maximizing the sum of the expected reward obtained at each time step. Next, we represent the state space, the action space and the reward function in our MDP as follows.

1) State Space: In a MDP the environment can be abstracted into a state space, which will change from one state to another after an action is selected by the agent. We define the state of MDP corresponding to the MEC system at time step $\tau$ as a vector $\textbf{s}_\tau=(\hat{\hat{B}},B_n^m,\hat{\hat{\delta}},w^r,w^m)$, which contains five parameters about computing delay and transmission delay. Specifically, $\hat{\hat{B}}$ denotes the sum of transmission rates between different ECDs in the network, i.e., $\hat{\hat{B}}=\sum\nolimits_{m=1}^{M}\sum\nolimits_{m'=1}^{M}B_{m,m'}$; $\hat{\hat{\delta}}$ expresses the sum of processing capacities of all ECDs in the network, i.e., $\hat{\hat{\delta}}=\sum\nolimits_{m=1}^{M}\delta_m$; $w^r$ indicates the total workloads of the tasks in the ready queue $Q^r$; $w^m$ is the total workloads of the tasks in computing queues of all ECDs. Note that, all parameters in $\textbf{s}_\tau$ can be observed by the agent from the real-time status information.

2) Action Space: The MDP uses the action to describe the behavior of an agent. Specifically, the action is the decision made by the agent after watching the state at a certain time step $\tau$. We use $\textbf{a}_\tau$ to denote the action space available to the agent when making scheduling decisions for each task. As each task except for the two dummy tasks can be scheduled to one of the ECDs in the network, $\textbf{a}_\tau=(a^0,a^1,\ldots, a^M)$ expresses the strategy space to schedule task $v_{ni}$. The item $a^m$ $(\forall m\in \{0,1,\ldots,M\})$ is the binary variable so that $\textbf{a}_\tau$ corresponds to the strategy $\textbf{x}_{ni}$ defined in Section \ref{Formulation}.

3) Reward: The MDP utilizes the reward to express the feedback from the environment to the agent. Specifically, the agent takes an action based on a decision, and then the environment provides feedback in the form of a reward value to the agent. In this work, the reward is the value that the agent calculates based on the state of the environment after it chooses an action for the ready task. The reward is defined as $\textbf{r}=\mathcal{U}-\mathcal{D}-\mathcal{P}$. Here $\mathcal{U}$ is a utility function, and $\mathcal{D}$ is a duration factor, and $\mathcal{P}$ expresses the penalty factor. We define $\mathcal{U}$ as follows.
\begin{equation}
	\label{U}
	\mathcal{U}=\beta\cdot\log_2{\rho_{ni}},
\end{equation}
where $\beta$ is a weight factor. $\mathcal{D}$ can be calculated by
\begin{equation}
	\label{D}
	\mathcal{D}=\psi\cdot(\mathop{\max}\limits_{t_{nj}\in p(t_{ni})}A(\textbf{x}_{nj})+Q(\textbf{x}_{ni})+E(\textbf{x}_{ni}))/\rho_{ni},
\end{equation}
where $\psi$ is a weight factor. Here, $A(\textbf{x}_{nj})$ can be obtained after $v_{ni}$'s each parent $v_{nj}$ is completed, as $v_{ni}$ is the ready task fetched from $Q^r$. Moreover, $\mathcal{P}$ can be defined as follows.
\begin{equation}
	\label{P}
	\mathcal{P}=\eta\cdot(F(\textbf{x}_{ni})-F_{ni}^{lct})/\rho_{ni},
\end{equation}
where $\eta$ is a weight factor.

\subsection{Deep Reinforcement Learning}
The reinforcement learning algorithm makes the decision based on the quality of actions, i.e., Q-value, when the agent interacts with the environment. However, in this work, the number of state spaces is so large that it is difficult to express them in the traditional Q-Table. Therefore, we introduce the DQN algorithm to solve the optimal scheduling decision. DQN uses the deep neural network to estimate the Q-value of the optimal decision, which can change Q-Table updating problem into the calculation of an approximation function. DQN can get similar output actions by the approximate value function after the agent observes the current state. Thus, the shortcoming of traditional Q-Table updating in high-dimensional and continuous problem can be overcome by a function fitting. Particularly, the approximate value function is represented as a parameterized functional form with parameters, as shown in the following formula.
\begin{equation}
	\begin{split}
		\label{Q_value}
		Q(\textbf{s}_\tau,\textbf{a}_\tau,\theta)\leftarrow & Q(\textbf{s}_\tau,\textbf{a}_\tau,\theta)\\&+\alpha(\textbf{r}_{\tau+1}+\gamma\max_{\hat{\textbf{a}}}Q(\textbf{s}_{\tau+1},\hat{\textbf{a}},\theta^{-})\\
		&-Q(\textbf{s}_{\tau},\textbf{a}_{\tau},\theta)),
	\end{split}
\end{equation}
where $\textbf{s}_{\tau+1}$ denotes the next state after the agent takes the action $\textbf{a}_{\tau}$ at time step $\tau$, $\textbf{r}_{\tau+1}$ is the reward that the agent receives at the next time step after taking action $\textbf{a}_{\tau}$, $\hat{\textbf{a}}$ is the action that maximizes the value at state $\textbf{s}_{\tau+1}$, $\gamma$ is the discount coefficient in the process of accumulating values, and $\alpha$ is the learning rate. Besides, $\theta$ and $\theta^{-}$ are parameter vectors at time step $\tau$ and $\tau+1$, respectively.

DQN algorithm can improve the search speed of the traditional Q-Learning by introducing the deep neural network, called the prediction network. The prediction network is responsible for estimating the Q-value at state $\textbf{s}_\tau$. The algorithm selects an action to make a scheduling decision. To improve the algorithm's stability and diversity, DQN sets the experience pool and adds the target network. The structure of the target network is consistent with that of the prediction network. The experience pool is the memory space that stores the records of the transition samples captured by the agent from the environment at each time step. Each record consists of four items ($\textbf{s}_\tau, \textbf{a}_\tau, \textbf{r}_{\tau+1}, \textbf{s}_{\tau+1}$), i.e., the state, the selected action, the received reward and the subsequent state. DQN extracts a batch of samples randomly from the experience pool to improve the data correlation and non-static distribution. Specifically, a batch of elements $\textbf{s}_\tau$ in the transition records is used as the input to the prediction network. Meanwhile, a batch of elements $\textbf{s}_{\tau+1}$ in the transition records is transmitted to the target network. During the training process, the parameters of the prediction network are copied to the target network after a certain number of round iterations. In DQN, the Q-value of the training process generated by the prediction network is called the current Q-value, and that of the training process generated by the target network is called the target Q-value. DQN maintains the difference between the parameters of the two neural networks to calculate the loss function. Whereafter, the parameters of the prediction network are inversely updated by leveraging the stochastic gradient descent (SGD). We define the loss function as follows.
\begin{equation}
	\begin{split}
		\label{Loss}
		L(\theta)=E[(Q_{tar}-Q_{cur})^2],
	\end{split}
\end{equation}
where $Q_{cur}=Q(\textbf{s}_\tau,\textbf{a}_\tau,\theta)$ is the current Q-value of the state-action pair outputted by the prediction network according to a batch of elements $\textbf{s}_\tau$ in the samples. $Q_{tar}$ is the target Q-value defined as follows.
\begin{equation}
	\begin{split}
		\label{Target_Q}
		Q_{tar}=\textbf{r}_{\tau+1}+\gamma\mathop{\max}_{\hat{\textbf{a}}}Q(\textbf{s}_{\tau+1},\hat{\textbf{a}},\theta^{-}),
	\end{split}
\end{equation}
where $Q(\textbf{s}_{\tau+1},\hat{\textbf{a}},\theta^{-})$ is the output of the target network according to a batch of elements $\textbf{s}_{\tau+1}$ in the samples.

\begin{figure}[htb]
	\centering
	\includegraphics[width=0.48\textwidth]{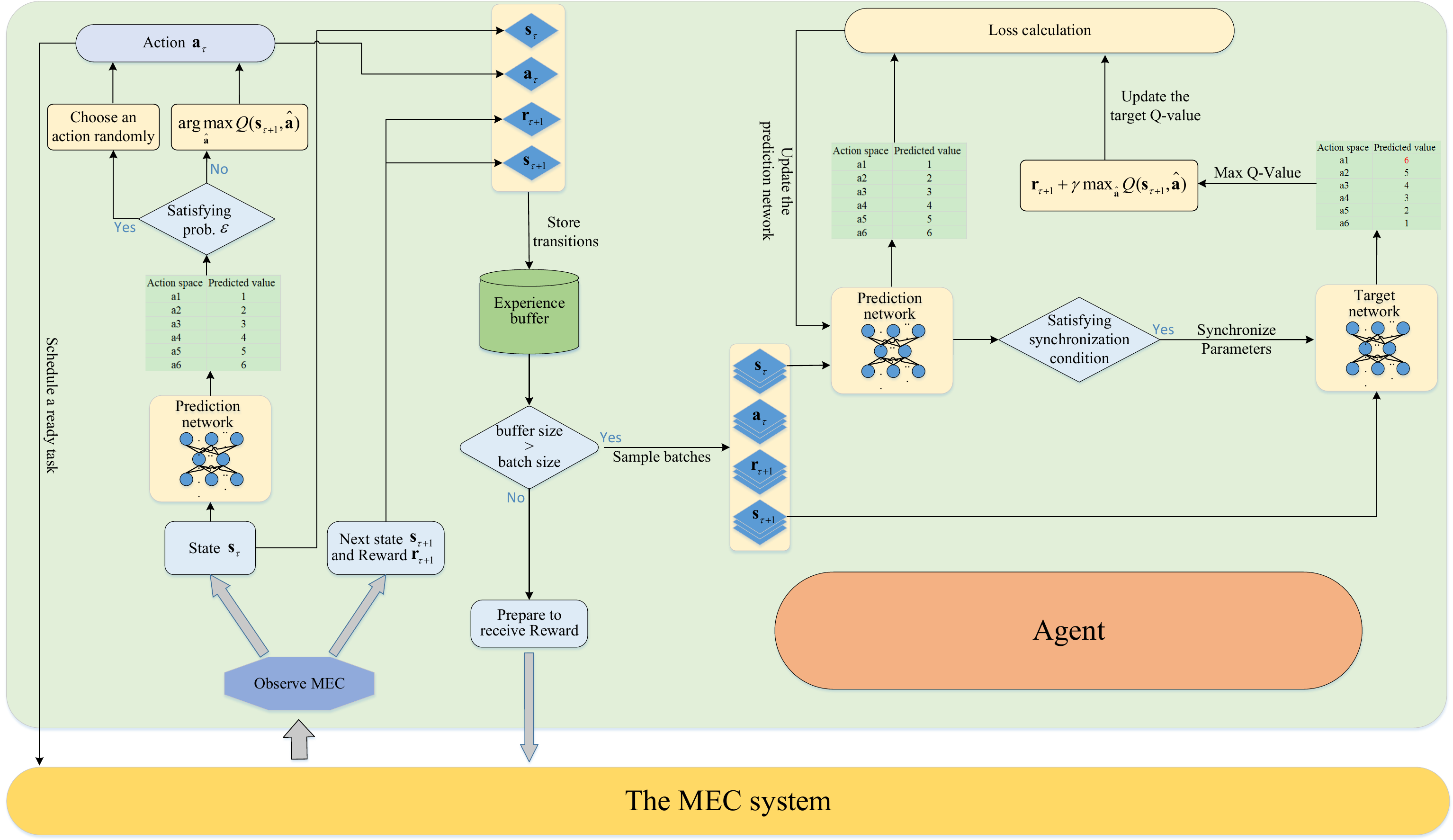}
	\caption{Framework of SATA-DRL.}
	\label{Framework}
\end{figure}

\section{Resource Allocation Based on Reinforcement Learning}
\label{ResourceAllocation}
We proposed a \underline{S}cheduling str\underline{A}tegy for the \underline{T}ask gr\underline{A}ph based on the \underline{D}eep \underline{R}einforcement \underline{L}earning, named as SATA-DRL, in order to solve the problem for computing resource allocation in the MEC system. To ease the understanding, we first present the reinforcement learning framework of scheduling strategy, and then introduce the scheduling algorithm for computing resource allocation.

\subsection{Reinforcement Learning Framework for Task Graph Scheduling}
The framework of SATA-DRL is shown in Fig.~\ref{Framework}. In this figure the agent is responsible for scheduling the ready tasks in the queue $Q^r$ by interacting with the environment of the MEC system. The agent has to be trained iteratively in the environment to determine optimized scheduling decisions. In training the agent first observes state $\textbf{s}_\tau$ from the environment at the time step $\tau$, and inputs $\textbf{s}_\tau$ into the prediction network. According to the Q-value output from the prediction network, an action $\textbf{a}_\tau$ is chosen for the task scheduling by using the $\varepsilon$-greedy exploration. The chosen action $\textbf{a}_\tau$ is defined as
\begin{equation}
	\begin{aligned}
		\label{exploration}
		\textbf{a}_\tau=\begin{cases} \text{an action chosen randomly},&\text{with probability}~\varepsilon,\\ 
			\mathop{\arg\max}\limits_{\hat{\textbf{a}}}Q(\textbf{s}_{\tau},\hat{\textbf{a}}),&\text{with probability}~1-\varepsilon.\end{cases}
	\end{aligned}
\end{equation}
Whereafter, the environment evolves to the next state $\textbf{s}_{\tau+1}$ and outputs the reward $\textbf{r}_{\tau+1}$. The agent observes $\textbf{s}_{\tau+1}$ and receives $\textbf{r}_{\tau+1}$ from the MEC system. The subsequent action is chosen based on the new Q-value after state $\textbf{s}_{\tau+1}$ is passed to the prediction network. At the same time, a transition record consisting of the state $\textbf{s}_{\tau+1}$, the reward $\textbf{r}_{\tau+1}$, the previous $\textbf{s}_\tau$ and the chosen $\textbf{a}_\tau$ is deposited into the experience pool. This process works repeatedly. After a few rounds of training, the agent randomly samples a batch of transitions from the experience pool. A batch of elements $\textbf{s}_\tau$ in these transitions is again transmitted to the prediction network as its inputs. Simultaneously, a batch of elements $\textbf{s}_{\tau+1}$ is used as the input to the target network. The agent chooses the maximum Q-value from the output from the target network, which is calculated as $\mathop{\max}_{\hat{\textbf{a}}}Q(\textbf{s}_{\tau+1},\hat{\textbf{a}})$. For example, the red value 6 in the output table from the target network in Fig~\ref{Framework} is the maximum Q-value. Then, based on Eq.~(\ref{Loss}), the agent calculates the loss function using the Q-value output from the prediction network and the Q-value output from the target network. Finally, the agent performs gradient descent with the goal of minimizing the variance-error for the parameter updating of the prediction network.

\subsection{Resource Allocation Algorithm for Task Graph}
{\color{blue}We design a resource scheduling algorithm, named SATA, for the task graph based on an event-driven mechanism. It can transform the information received from the environment into the state and reward in the MDP model. Then, the state and reward will be passed to the deep reinforcement learning algorithm, called DRL. SATA takes charge of fetching out all the ready tasks from $\mathcal{H}$ and pushing them into queue $Q^r$ to invoke DRL.}


Algorithm~\ref{SATA} shows the processing of the task graph scheduling. This algorithm is launched on receipt of the arrival and completion events. In Lines 2-7 the tasks of an arrived application $n$ are sorted in list $\xi_n$ by an ascending order of $F_{ni}^{lct}$, meanwhile, and list $\xi_n$ is appended to $\mathcal{H}$ when the arrival event triggers. This algorithm's Lines 8-19 will be launched when either the arrival event or the completion event triggers. Specifically, Lines 8-11 express that all ready tasks in the set $\mathcal{H}$ are mapped into $Q^r$ by an ascending order of the latest completion time. In Lines 12-19 SATA generates the sequence of discrete time steps for the task graph scheduling by iteratively fetching out a ready task $v_{ni}$ from $Q^r$. SATA obtains the current state $\textbf{s}_\tau$ from the environment at each time step $\tau$. Notice that, if the fetched task is the first of all application tasks, the reward is a null value. Otherwise, SATA will calculate the current reward $\textbf{r}_\tau$ based on Eqs.~(\ref{U}),(\ref{D}),(\ref{P}) according to the environment information. In Line 17 SATA passes $\textbf{s}_\tau$ and $\textbf{r}_\tau$ to the deep reinforcement learning algorithm, and subsequently receives the action space $\textbf{a}_\tau$ from it. Thus, the task $v_{ni}$ can be scheduled to the target ECD in the light of $\textbf{a}_\tau$. The SATA algorithm will end when no new applications arrive or no tasks are completed.

\begin{algorithm}[htb]
	\caption{SATA (receipt of two events)}
	\label{SATA}
	\begin{algorithmic}[1]
		\STATE{$\mathbb{O}\gets\varnothing$; ${Q}^{r}\gets\varnothing$; $\tau\gets 1$; $\textbf{r}\gets\varnothing$;}
		\IF{ A new application $n$ arrives at the MEC system}
		\STATE{Add application $n$ into $\mathbb{O}$;}
		\FOR{each application $n\in\mathbb{O}$}
		\STATE{Estimate $F_{ni}^{lct}$ for each $v_{ni}$ based on Eq.~(\ref{LCT});}
		\STATE{Rank all tasks in list $\xi_n$ by an ascending order of $F_{ni}^{lct}$;}
		\STATE{ Append $\xi_n$ to $\mathcal{H}$;}
		\ENDFOR
		\ENDIF
		\FOR{each $\xi_n\in\mathcal{H}$}
		\IF{$v_{ni}$ in the head of $\xi_n$ is the ready task}
		\STATE{Map $v_{ni}$ into ${Q}^{r}$;}
		\ENDIF
		\ENDFOR
		\STATE{Rank ${Q}^{r}$ by an ascending order of $F_{ni}^{lct}$;}
		\FOR{each $v_{ni}$ in ${Q}^{r}$}
			\STATE{Obtain state $\textbf{s}_\tau$ from the environment at step $\tau$;}
			\IF{$\tau>1$}
			\STATE{Calculate $\mathcal{U}$, $\mathcal{D}$ and $\mathcal{P}$ based on Eqs.~(\ref{U}),(\ref{D}),(\ref{P});}
			\STATE{$\textbf{r}\gets\mathcal{U}-\mathcal{D}-\mathcal{P}$; }
			\ENDIF
			\STATE{Pass $\textbf{s}_\tau$ and $\textbf{r}$ to the DRL algorithm, and then receive $\textbf{a}_\tau$ from the DRL algorithm;}
			\STATE{Schedule $v_{ni}$ to the target ECD based on $\textbf{a}_\tau$;}
			\STATE{$\tau\gets\tau+1$;}
		\ENDFOR
		\end{algorithmic}
	\end{algorithm}
	
For the DRL algorithm, it is to implement deep reinforcement learning. Algorithm~\ref{DRL} shows the processing that SATA makes the scheduling decision based on deep reinforcement learning. This algorithm first constructs two neural networks with stochastic parameters $\theta$ and $\theta^-$, respectively. Then, it will receive state $\mathcal{S}$ and reward $\mathcal{R}$ from the SATA algorithm at each time step. The prediction network outputs the Q-value by inputting state $\mathcal{S}$. Based on the $\varepsilon$-greedy exploration, action $\mathcal{A}$ is chosen out. The above processes are described in lines 4-7 of this algorithm. Notice that, if the current time step is processing  the first task of all applications to be scheduled, i.e., $\tau=1$, the algorithm will skip the storage of the experience pool and the computation of the target network, and it will prepare the state and action for storing the experience transition in the next time step. So, Lines 8-16 are skipped and Line 17 is run directly when $\tau=1$. When $\tau>1$, the transition consisting of the last state $\textbf{s}_\tau$ and action $\textbf{a}_\tau$ as well as the current $\textbf{r}_{\tau+1}$ and $\textbf{s}_{\tau+1}$ is stored into the experience pool $\mathbb{D}$, as shown in Line 10. Then, the DRL algorithm will sample a batch $\mathcal{B}$ of transitions from $\mathbb{D}$ when the number of time steps is more than the predefined batch size, as shown in Lines 11-12. For each transition in $\mathcal{B}$, the DRL algorithm needs to calculate the loss function based on the output from the prediction network and target network, respectively, and it finally performs SGD to update the prediction network. These pseudo-codes are shown in Lines 13-16.

	\begin{algorithm}[htb]
		\caption{DRL algorithm}
		\label{DRL}
		\begin{algorithmic}[1]
			\REQUIRE {The step size $K$ for updating the target network; The batch size $batch$ for sampling transitions;}
			\STATE{Initialize the prediction network and target network with stochastic parameters $\theta$ and $\theta^-$, respectively;}
			\STATE{$\mathbb{D}\gets\varnothing$;}
			 \FOR{$k=1, 2,\cdots, K$}
			 	\STATE{Receive state $\mathcal{S}$ and reward $\mathcal{R}$ from the SATA algorithm at step $\tau$;}
			 	\STATE{Get the Q-value in the prediction network using $\mathcal{S}$;}
			 	\STATE{Choose action $\mathcal{A}$ by utilizing Eq.~(\ref{exploration});}
			 	\STATE{Pass $\mathcal{A}$ to the SATA algorithm;}
			 	\IF{$\tau>1$}
			 	\STATE{$\textbf{s}_{\tau+1}\gets \mathcal{S}$; $\textbf{r}_{\tau+1}\gets\mathcal{R}$;}
			 	\STATE{Store transition ($\textbf{s}_{\tau}$, $\textbf{a}_\tau$, $\textbf{r}_{\tau+1}$, $\textbf{s}_{\tau+1}$) into $\mathbb{D}$;}
			 	\IF{$k> batch$}
			 	\STATE{Sample a mini-batch $\mathcal{B}$ of transitions from $\mathbb{D}$;}
			 	\FOR{Each transition ($\textbf{s}_{i}$, $\textbf{a}_i$, $\textbf{r}_{i+1}$, $\textbf{s}_{i+1}$) in $\mathcal{B}$}
			 	\STATE{Get $Q_{cur}$ in the prediction network using $\textbf{s}_{i}$;}
			 	\STATE{Get $Q(\textbf{s}_{i+1},\hat{\textbf{a}},\theta^{-})$ in the target network using $\textbf{s}_{i+1}$, and then calculate $Q_{tar}$ according to Eq.~(\ref{Target_Q});}
			 	\STATE{Perform SGD to update the prediction network;}
			 	\ENDFOR
			 	\ENDIF
			 	\ENDIF
			 	\STATE{$\textbf{s}_{\tau}\gets \mathcal{S}$; $\textbf{a}_\tau\gets \mathcal{A}$;}
			 \ENDFOR
			 \STATE{$\theta^-\gets\theta$;}
		\end{algorithmic}
	\end{algorithm}

\subsection{Intelligent Agent Training and Working}
Once a reinforcement learning algorithm has been designed, the agent needs to be trained to learn the intelligence. Algorithm~\ref{Train} represents the process of agent training. $\Omega$ is the number of episodes for agent training. We will preset a value of $\Omega$ to try to achieve network convergence. 

\begin{algorithm}[htb]
	\caption{Agent training}
	\label{Train}
	\begin{algorithmic}[1]
		\REQUIRE{The maximum number $\Omega$ of episodes}
		\FOR{$e=1, 2,\cdots, \Omega$}
		\STATE{Invoke the DRL algorithm;}
		\ENDFOR
	\end{algorithmic}
\end{algorithm}

The agent can work alone for the task graph applications in the MEC system after deep neural networks converge. At this moment, the agent can obtain the optimized scheduling decision by the output from the prediction network according to state $\textbf{s}_\tau$ in each time step. Algorithm~\ref{Work} shows that the agent makes the task scheduling decision by the output from the prediction network. This algorithm is similar to Algorithm~\ref{SATA} in that it also starts working based on the arrival and completion events. Since the neural network for reinforcement learning has converged, the prediction network can directly output $Q(\textbf{s}_{\tau},\hat{\textbf{a}})$  according to the state $\textbf{s}_\tau$ obtained from the environment, as shown in Lines 13-14. Then, Algorithm~\ref{Work} chooses the action $\textbf{a}_\tau$ that maximizes $Q(\textbf{s}_{\tau},\hat{\textbf{a}})$ for the task scheduling.

\begin{algorithm}[htb]
	\caption{SATA-Work (recept of two events)}
	\label{Work}
	\begin{algorithmic}[1]
		\REQUIRE{the set  $\mathcal{N}$ of all applications to offload}
		\STATE{$\mathbb{O}\gets\varnothing$; ${Q}^{r}\gets\varnothing$;}
		\IF{ A new application $n$ arrives at the MEC system}
		\STATE{Add application $n$ into $\mathbb{O}$;}
		\FOR{each application $n\in\mathbb{O}$}
		\STATE{Estimate $F_{ni}^{lct}$ for each $v_{ni}$ based on Eq.~(\ref{LCT});}
		\STATE{Rank all tasks in list $\xi_n$ by an ascending order of $F_{ni}^{lct}$;}
		\STATE{ Append $\xi_n$ to $\mathcal{H}$;}
		\ENDFOR
		\ENDIF
		\FOR{each $\xi_n\in\mathcal{H}$}
		\IF{$v_{ni}$ in the head of $\xi_n$ is the ready task}
		\STATE{Map $v_{ni}$ into ${Q}^{r}$;}
		\ENDIF
		\ENDFOR
		\STATE{Rank ${Q}^{r}$ by an ascending order of $F_{ni}^{lct}$;}
		\FOR{each $v_{ni}$ in ${Q}^{r}$}
		\STATE{Obtain state $\textbf{s}_\tau$ from the environment;}
		\STATE{Get $\textbf{a}_\tau=\mathop{\arg\max}\nolimits_{\hat{\textbf{a}}}Q(\textbf{s}_{\tau},\hat{\textbf{a}})$ in the prediction network using $\textbf{s}_\tau$;}
		\STATE{Schedule $v_{ni}$ to the target ECD based on $\textbf{a}_\tau$;}
		\ENDFOR
	\end{algorithmic}
\end{algorithm}

\begin{figure*}[hbp]
	\centering
	\subfigure[$\lambda=5$]{
		\label{Lambda5} 
		\includegraphics[width=5.5cm]{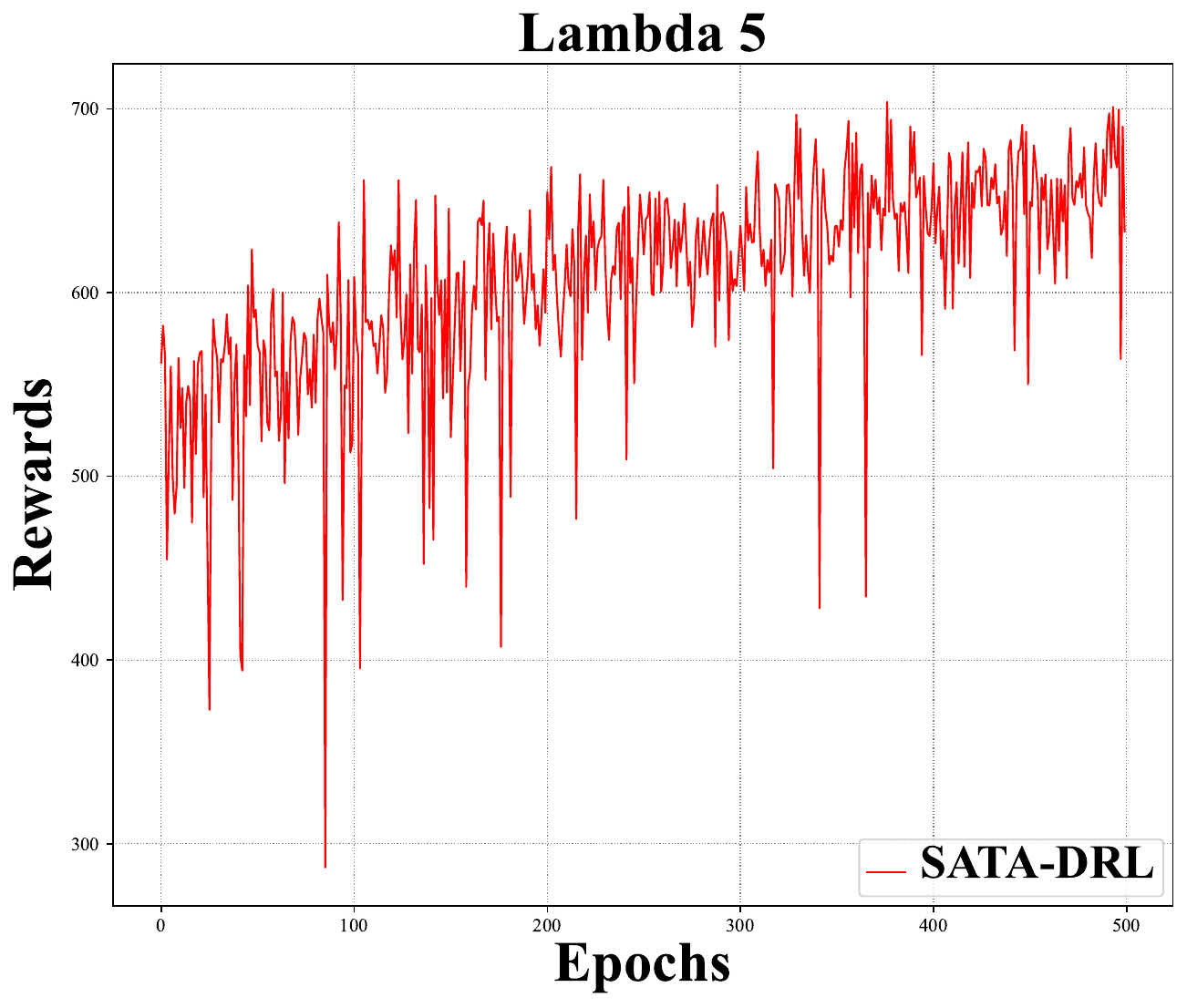}}
	\hspace{0in}
	\subfigure[$\lambda=7$]{
		\label{Lambda7} 
		\includegraphics[width=5.5cm]{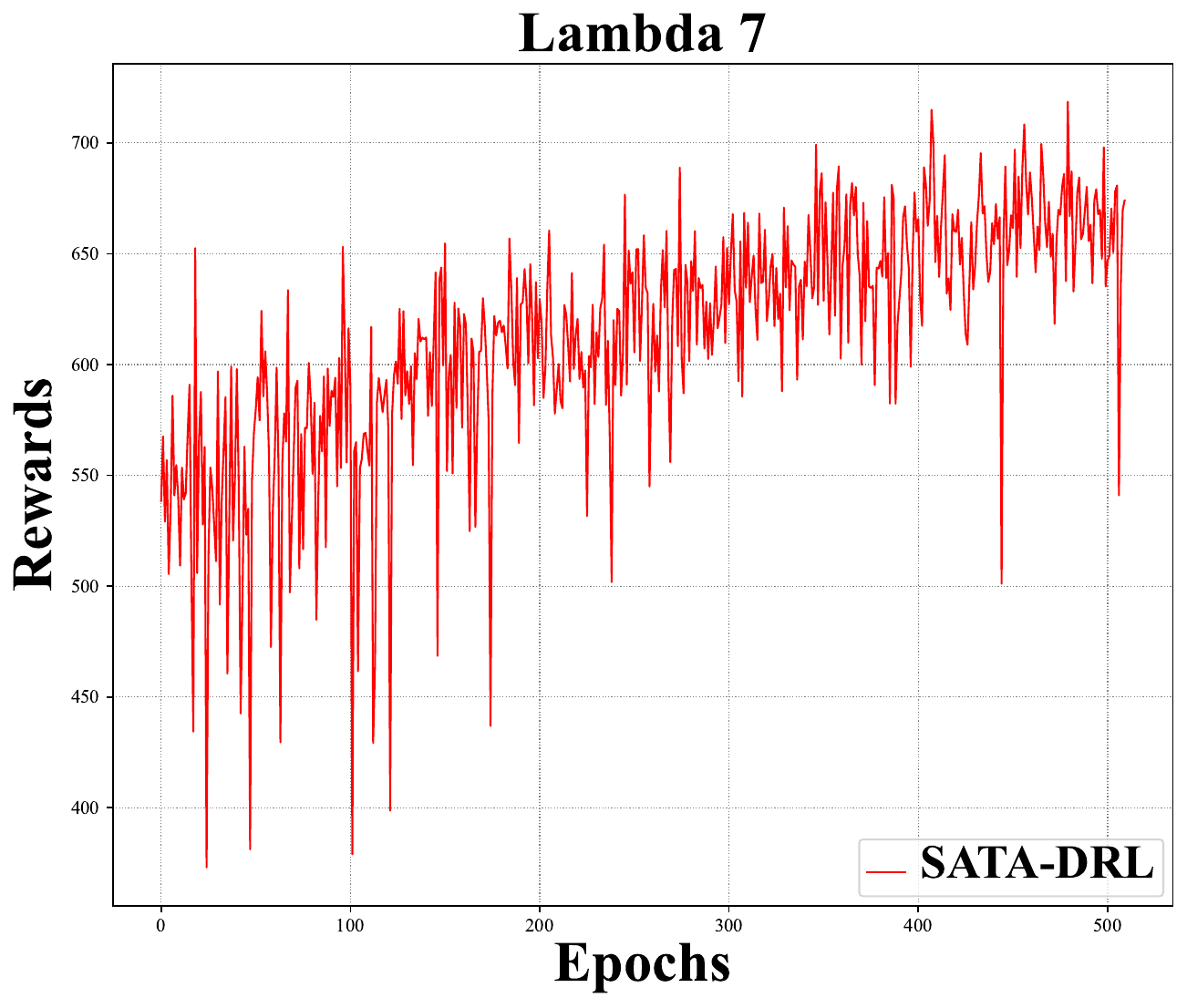}}
	\hspace{0in}
	\subfigure[$\lambda=9$]{
		\label{Lambda9} 
		\includegraphics[width=5.5cm]{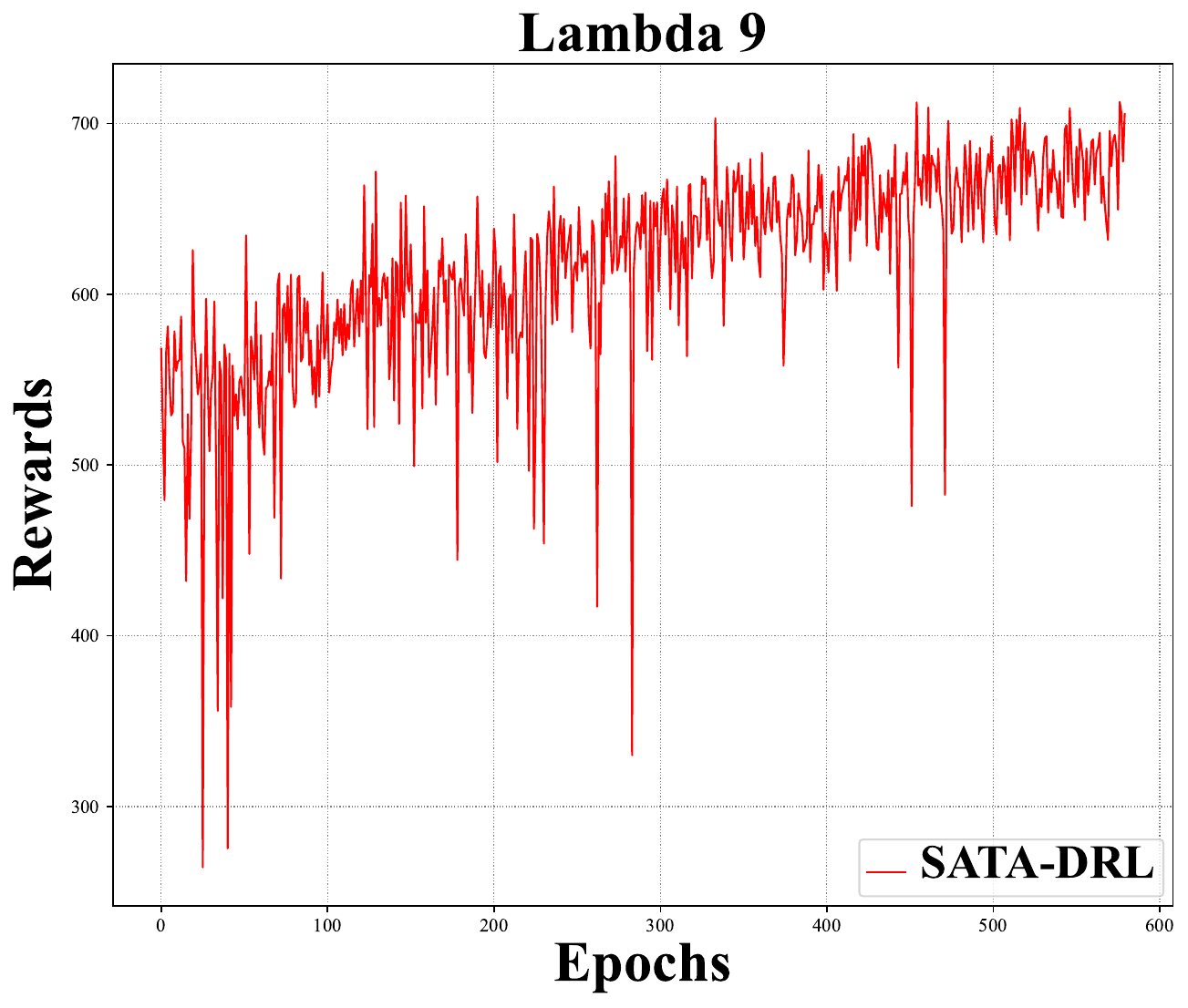}}
	\caption{Cumulative rewards with different $\lambda$}
	\label{Cumulative_reward} 
\end{figure*}

\section{Performance evaluation}
\label{Performance}
{\color{blue}To evaluate SATA-DRL, we extend CloudSim~\cite{calheiros2011cloudsim} according to two simulators, i.e., EdgeCloudSim~\cite{sonmez2018edgecloudsim} and ElasticSim~\cite{cai2017elasticsim}. Specifically, we tailor CloudSim to the problem of dependent task offloading by combining two approaches as follows. The one is the simulation of edge service devices supported by the EdgeCloudSim simulator. The other is the generation of workflows supported by the ElasticSim simulator.} Moreover, we compare SATA-DRL with several competing algorithms for the evaluation of task graph scheduling in the MEC system.

\subsection{Simulation Setup}
There are four ECDs in our simulator, and the processing capability of each MU, i.e., the $0$-th edge computing device, is set to 1000 MIPS. In practice, the processing capability of each ECD's PE may change over time. Similar to the literature~\cite{ning2019deep}, we divide the processing capability of each ECD's PE into discrete levels: $\{6000, 5500, 5000, 4500, 4000\}$ (MIPS). Each level corresponds to one processing capability state. Thus, the transition of the processing capability level of each ECD can be modeled as a Markov chain, whose state transition probability matrix can be derived as follows~\cite{ning2019deep}.
\begin{equation}
	\begin{bmatrix}
		0.5& 0.25& 0.125& 0.0625& 0.0625\\
		0.0625& 0.5& 0.25& 0.125& 0.0625\\
		0.0625& 0.0625& 0.5& 0.25& 0.125\\
		0.125& 0.0625& 0.0625& 0.5&0.25\\
		0.25& 0.125& 0.0625& 0.0625& 0.5
	\end{bmatrix}
\end{equation}
In our experiments, we let the processing capabilities of four ECDs transition from one state to another according to this probability matrix after a task is completed on any computing device. Besides, the transmission rate between ECDs is set to 440 Mbps, and that between each MU and ECD $m_n$, i.e., the ECD covering the MU, is set to $10^3$  Mbps. These two transmission rates are similar to the literature~\cite{sundar2018offloading}. In addition, $\hat{\delta}$ and $\hat{B}$ are set to 6000 MIPS and $10^3$  Mbps, respectively.

{\color{blue}The scientific workflow datasets~\cite{juve2013characterizing} contain the workflow characterizations corresponding to task graph applications. It can provide a community resource to evaluate workflow algorithms. As Montage workflow has been used extensively to evaluate workflow algorithms and systems, the literature~\cite{zhang2018resource} utilized the Montage workflow to evaluate its task graph offloading algorithm. Similar to~\cite{zhang2018resource}, We choose the Montage type of workflow datasets, which contains 25 nodes, to simulate the task graph of each UT's application. }
The workload of a task in an application is set to 500 if the original value of the element $runtime$ recorded in the DAX file is greater than 500. Otherwise, that of a task is set to 100 if the original $runtime$ is less than 100. To set the  size of transfer data, we need to calculate the average transmission rate $\overline{B}$ as $(440\times 6+10^3)/7=520$ (Mbps) in the MEC system. Then, based on the element $size$ recorded in the DAX file, we can calculate a base communication time, named $bc$. We set $bc$ to $size/\overline{B}$, if $bc=size/\overline{B}$ does not exceed the range of $[10^{-3},10^{-2}]$. Otherwise, let $bc$ equal to $10^{-3}$ and $10^{-2}$, respectively. Based on this, we set the size of the data transfer between tasks in an application to $bc\cdot\overline{B}$. The size of the data transfer associated with the dummy tasks is set to $bc\cdot\overline{B}$.

Since the offloading request of MUs can randomly generate, we fetch some applications from Montage workflows with a stochastic time interval, following the Poisson distribution with $\lambda$. Furthermore, to simulate the stochastic arrival of applications in the MEC system, we let MUs randomly generate the offloading request in the celluar communication service area of four ECDs, following a uniform distribution. In addition, to set the deadline of each application, we assume that each task in a fetched application $n$ is scheduled to the different ECD with the maximum average processing capability, i.e., $5000=(6000\times 4+1000)/6$, among the MEC system, and all transfer data among tasks in application $n$ are ignored. Thus, we can calculate a basic makespan $MS_n$ for each application $n$. Whereafter, we set the deadline of application $n$ to $d_n=r_n+6\cdot MS_n$.

\begin{figure*}[htp]
	\centering
	\subfigure[$\lambda=5$]{
		\label{Make5} 
		\includegraphics[width=5.5cm]{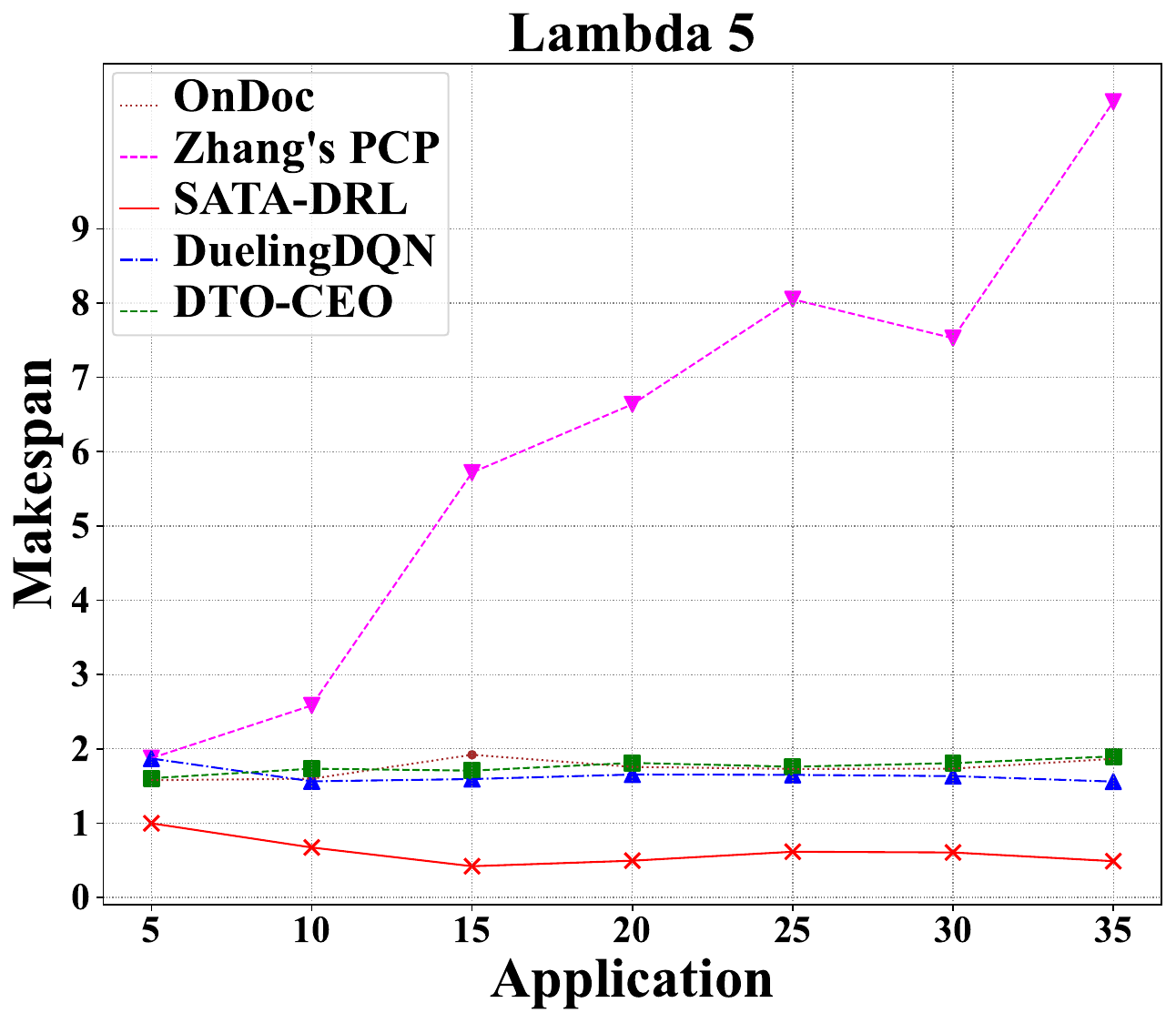}}
	\hspace{0in}
	\subfigure[$\lambda=7$]{
		\label{Make7} 
		\includegraphics[width=5.5cm]{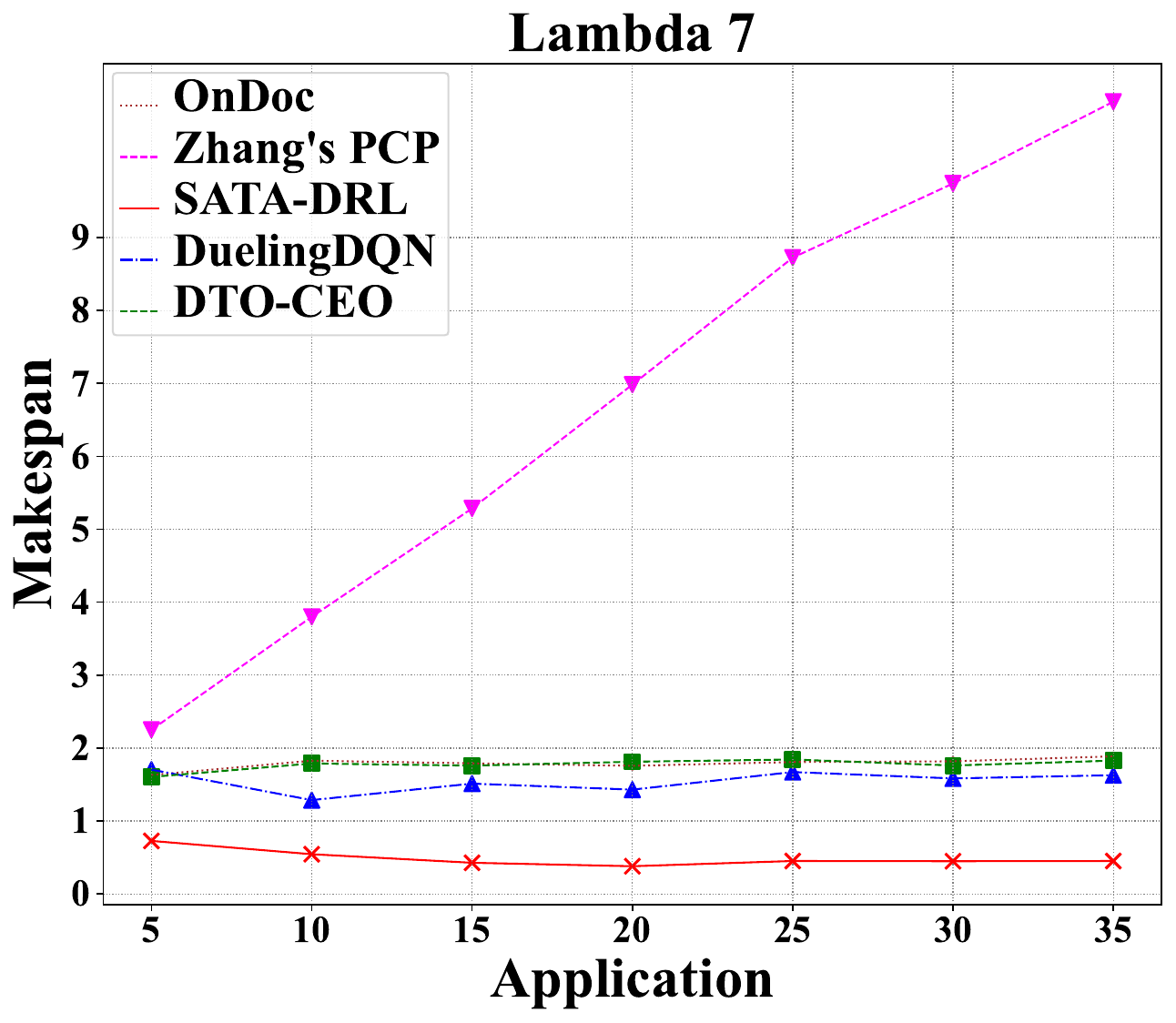}}
	\hspace{0in}
	\subfigure[$\lambda=9$]{
		\label{Make9} 
		\includegraphics[width=5.5cm]{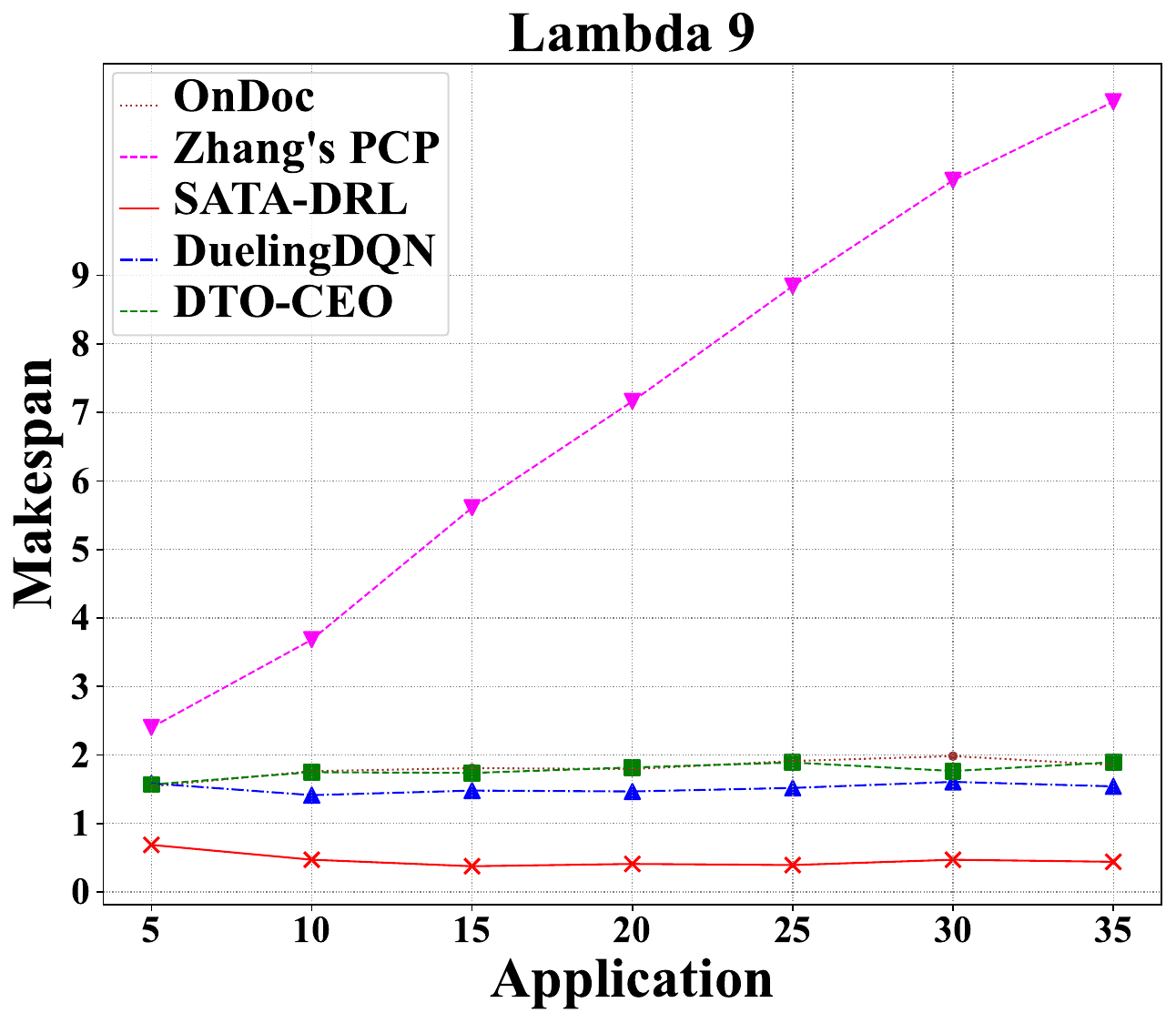}}
	\caption{Average Makespan Comparison}
	\label{MakespanComparison} 
\end{figure*}

\subsection{Setup for Intelligent Agent}
There are two neural networks with the same structure in SATA-DRL. In our experiments, these neural networks are fully connected networks consisting of five layers. The number of nodes in these networks are set to 128, 64, 32, 16, 5, respectively. The activation function is $linear$ in the first four layers, and that is $softmax$ in the last layers. We use $Adam$ as the optimization method for the gradient descent and set the learning rate of the agent to 0.0006. 

Moreover, we set the size of the experience pool for DQN to 200000. In Algorithm~\ref{DRL} the batch size $batch$ is set to 64. $\beta$ of Eq.~(\ref{U}) is set to 0.6 and $\psi$ of Eq.~(\ref{D}) is set to 5. Meanwhile, $\eta$ of Eq.~(\ref{P}) is set to 40. Besides, in Eq.~(\ref{Target_Q}), we set the discount factor $\gamma$ for the calculation of the target Q-value to 0.95.

\subsection{Evaluation for Reinforcement Learning}
In this section, we use cumulative rewards to evaluate the performance of our reinforcement learning.
Considering that MUs can randomly send offloading requests of applications, we evaluate cumulative rewards of the reinforcement learning with different arrival rates, i.e., $\lambda=\{5,7,9\}$, when it handles 10 applications. In the experiments, we let 10 applications per episode reach the simulator with different arrival rates, and then launch Algorithm~\ref{Train} to train the agent and observe the reward values. From Figs.~\ref{Cumulative_reward}, we can observe that the total rewards obtained by the agent are not large at the beginning of the learning process. The total rewards increase as we augment the number of episodes for training. Moreover, the total rewards show relatively stable trends when agent is trained for more than 600 episodes, which is shown in Fig.~\ref{Lambda9}. This can be interpreted that the agent can make a relatively optimized decision in every scheduling for the task. As a result, the cumulative rewards can converge. The above results demonstrate the convergence performance of SATA-DRL.

\begin{figure*}[htp]
	\centering
	\subfigure[$\lambda=5$]{
		\label{Viol5} 
		\includegraphics[width=5.5cm]{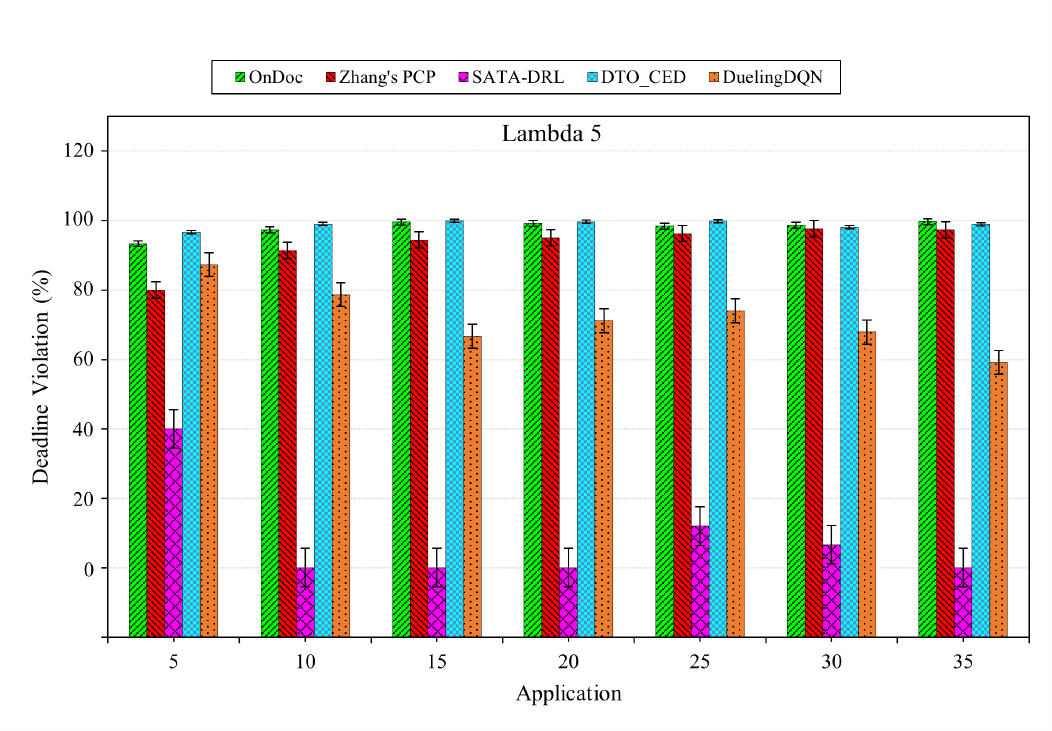}}
	\hspace{0in}
	\subfigure[$\lambda=7$]{
		\label{Viol7} 
		\includegraphics[width=5.5cm]{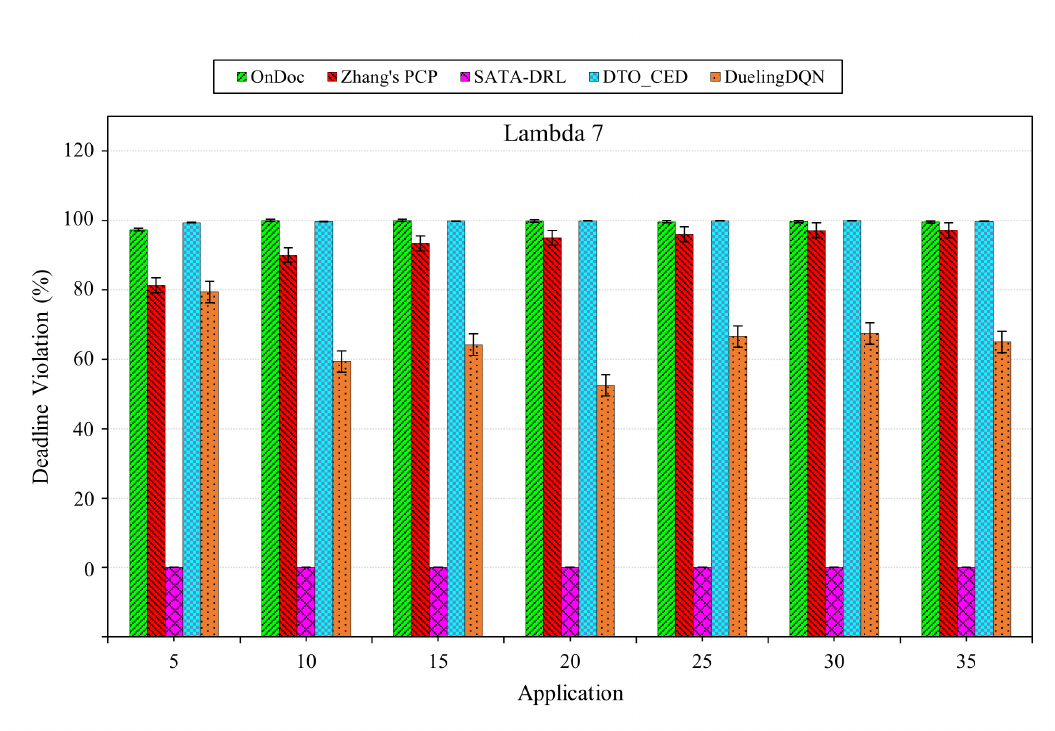}}
	\hspace{0in}
	\subfigure[$\lambda=9$]{
		\label{Viol9} 
		\includegraphics[width=5.5cm]{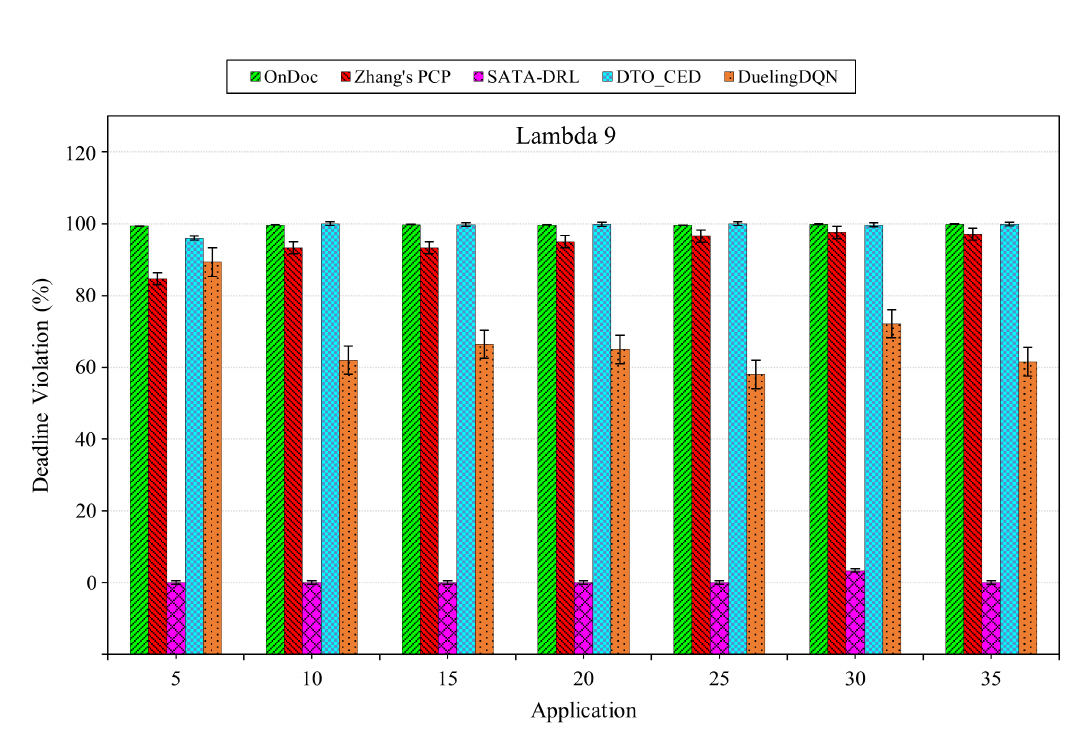}}
	\caption{Deadline Violation Comparison}
	\label{Violation} 
\end{figure*}

\subsection{Evaluation for Optimization Objective}
\textcolor{blue}{STATA-DRL is a task graph offloading strategy for multiple applications in the MEC system to minimize the average makespan of applications. There is no fine-grained task graph offloading that utilize DRL to learn the time-varying computing capabilities of ECDs. To evaluate the optimization performance of SATA-DRL, we compare it with three competitors based on heuristic or approximation methods, i.e., Zhang's PCP\cite{zhang2018resource}, OnDoc\cite{Liuyan2019online}, and DTO-CED~\cite{zhang2023dependent}, which are also to minimize the makespan of task graph applications MEC system. Moreover, we compare SATA-DRL with a new reinforcement algorithm, named Dueling DQN, to advantages of SATA-DRL. In the experiments that evaluate Dueling DQN, we still utilize the MDP model presented in Section 5.2, that is, the state space, the action space, and the reward are the same as SATA-DRL. The Dueling DQN receives the input of the state space and reward and then outputs the action.} 

In these group experiments, let the three algorithms handle many applications with different arrival rates. We randomly fetch many applications following Possion distribution with $\lambda$ to store them into a file. Each of four algorithms loads the fetched applications from the file for performance evaluation. To alleviate the randomness of the experimental results, the four algorithms handle the same applications for 30 times, respectively. Then, we utilize the averages of results to plot. In experiments, we evaluate the average makespan and the deadline violation rate for all applications, respectively. The deadline violation rate can reveal how many applications will lose their deadlines. We express the total number of applications as $|\mathcal{N}|$ and indicate the total number of applications violating deadlines as $|\overrightarrow{\mathcal{N}}|$. Thus, the deadline violation is defined as $|\overrightarrow{\mathcal{N}}|/|\mathcal{N}|\times100\%$.

Fig.~\ref{MakespanComparison} shows the makespan comparison for handling various applications with $\lambda=\{5,7,9\}$, respectively. {\color{blue}We can observe that SATA-DRL achieves the lowest makespan among the five algorithms. Particularly, the makespans generated by Zhang's PCP increase along with the increase of application arrival rates but the other algorithms are hardly affected by arrival rates. This is because that Zhang's PCP only make scheduling decision for the task among a single application every time without considering the arrvial of multiple applications. Therefore, the application tasks newly sent from other MUs may still be scheduled to the busy ECDs, resulting in backlogs in the computing queues of ECDs. In addition, the gap among SATA-DRL, DTO-CEO, and OnDoc is relatively small but the makespan achieved by SATA-DRL is less than DTO-CEO and OnDoc.} The reason is that SATA-DRL can learn from the variation of environments and adaptively make more appropriate scheduling decisions for task graph scheduling. Furthermore, we can observe from Fig.~\ref{Violation} that the deadline violation rate of SATA-DRL is much lower than the other algorithms. This further illustrates that SATA-DRL can fully orchestrate the computing resources of the system to make scheduling decisions with careful consideration of the variation of environments.

{\color{blue}From Figs.~\ref{MakespanComparison} and \ref{Violation}, we can also observe that SATA-DRL is superior to Dueling DQN in terms of reducing average makespan and deadline violation. The comparison with the other reinforcement algorithm confirms that the classic DQN method achieves better performance of task graph offloading compared to Dueling DQN.}

\section{Conclusions}
\label{Conclusion}
This paper has investigated the problem of the task graph offloading in MEC, where the computation capabilities of edge computing devices are time-varying. To adapt to environmental changes, we have modeled the task graph scheduling for computation offloading as a MDP. According to the characterization of the environment, we have formulated it as the state space and abstracted the task scheduling decisions into the action space. Moreover, we have defined 
the reward with respect to MDP as the benefit for the agent. Built upon the task graph scheduling mechanism, we have designed the SATA-DRL algorithm to learn the task scheduling strategy from the interaction with the environment, improving user experience. Extensive experiment results have also validated the superiority of SATA-DRL, by comparing it with existing algorithms, in terms of reducing average makespan and deadline violation.

\section*{Acknowledgments}
This work is supported in part by Hunan Provincial Natural Science Foundation of China under Grant 2022JJ50147, in part by the 14-th Five-Year-Plan Project of Hunan Provincial Education Science under Grant ND228128.



\bibliographystyle{elsarticle-num} 
\bibliography{reference}






\end{document}